\documentclass[12pt,letterpaper]{JHEP3}

\usepackage{amsmath}
\usepackage{amssymb}
\usepackage{amsfonts}
\usepackage{graphicx}

\DeclareSymbolFont{AMSa}{U}{msa}{m}{n}
\DeclareSymbolFont{AMSb}{U}{msb}{m}{n}
\let\Box\relax
\DeclareMathSymbol{\Box}{\mathord}{AMSa}{"03}

\title{Microscopic Quantum Mechanics of the $p=\rho$ Universe}

\author{T. Banks \\
Department of Physics and Astronomy - NHETC\\
Rutgers University, Piscataway, NJ 08540\\
and\\
Department of Physics, SCIPP\\
University of California, Santa Cruz, CA 95064\\
E-mail: \email{banks@scipp.ucsc.edu}}

\author{W. Fischler, \\
Department of Physics\\
University of Texas, Austin, TX 78712\\
E-mail: \email{fischler@utexas.edu}}

\author{L. Mannelli \\
Department of Physics, SCIPP\\
University of California, Santa Cruz, CA 95064\\
E-mail: \email{lorenzo@scipp.ucsc.edu}}

\abstract{We present a complete quantum mechanical description of a flat
FRW universe with equation of state $p=\rho$. We find a detailed
correspondence with our heuristic picture of such a universe as a dense
black hole fluid. Features of the geometry are derived from purely quantum
input.}


\preprint{\hepth{} \\SCIPP-04/12\\ UTTG-08-04}

\begin{document}



\section{Introduction}

A little over two years ago, two of us (TB and WF) introduced a new approach
to cosmological initial conditions called holographic cosmology \cite%
{holocosmo}. The basic principle on which it was based is the holographic
entropy bound \cite{fsb}\cite{g}. In a Big Bang cosmology, the bound
implies a finite entropy for any causal diamond\footnote{%
In fact, all of our previous work referred instead to the causal past of a
point. Raphael Bousso has repeatedly emphasized the greater virtues of
causal diamonds (where every point can be both seen and influenced by an
observer) and we have realized that all of our actual formulae could be
taken to refer to causal diamonds rather than causal pasts.} whose future
boundary is a finite timelike separation from the Big Bang. This entropy
decreases to zero as we approach the initial singularity. We interpreted
this entropy as the entropy of the maximally uncertain density matrix for
measurements done inside the causal diamond, a conjecture with several
attractive features.

Our approach led us both to a tentative set of rules for defining a general
quantum space-time, and to a heuristic approach to the Big Bang singularity.
In this paper we close the circle of these ideas. We find a solution of the
consistency conditions we have formulated for quantum cosmology, which
behaves qualitatively like the dense black hole fluid which was the basis
for our heuristic description.

The mathematical formalism which we will present in this paper was alluded
to in several of our previous publications \cite{susyholo3.0}. It is
motivated by the results of Belinskii, Khalatnikov and Lifshitz (BKL) and
subsequent workers, which suggest that dynamics near a Big Bang
singularity is chaotic \cite{BKL}. This leads us to postulate that the
time dependent Hamiltonian near the Big Bang, is, at each instant chosen
independently from a certain random distribution of Hamiltonians. We will
describe the distribution in more detail in section 3. For large causal
diamonds, this hypothesis leads to a time independent spectral density for
the time dependent Hamiltonian; that of a $1+1$ dimensional conformal
field theory. Thus, the system is given a random kick at each time, but
the spectral density of the time dependent Hamiltonian approaches a
universal limit. The energy/entropy density relation $\sigma \sim
\sqrt{\rho }$ of this system is precisely that of our heuristic black hole
fluid, and is the relation following from thermodynamics and extensivity
in any dimension, for a fluid with equation of state $p=\rho $.

Guided by this correspondence, we argue that the energy per unit length of
the $1+1$ dimensional system should be taken as the space time Hamiltonian
for an observer in a given causal diamond in the $p=\rho$ background. Using
the transformation between entropy and cosmological time, we show that this
observer, in most of the states of the $1+1$ dimensional system, sees an
energy precisely equal to the mass of a horizon filling black hole.

We then show that the basic structure of our quantum formalism allows us to
derive the $d$ dimensional space time metric, which is a flat FRW universe
with perfect fluid matter satisfying the equation of state $p=\rho $. The
scaling symmetry of the $1+1$ CFT is reinterpreted as invariance of the
dynamics under the conformal Killing vector of this cosmology. This symmetry
was crucial to our derivation \cite{holcosm1and2} of a scale invariant
fluctuation spectrum for the cosmic microwave background.

We have structured this paper in the following manner: In the next section
we present a general framework for the local quantum dynamics of
gravitational systems. The formalism associates operator algebras with
causal diamonds in a space time. The details of the mapping depend on the
nature of the boundaries of space time. The fundamental quantum variables
are associated with holographic screens for a causal diamond by the
Cartan-Penrose \cite{cp} equation. Heuristically, we view them as
\textquotedblleft quantum pixels on the holographic screen of a causal
diamond". They transform as spinors under local Lorentz transformations and
inherit a natural $Z_{2}$ gauge invariance from the classical CP equation.
We use this gauge symmetry to transform them into fermions, explaining the
conventional connection between spin and statistics.

In Section 3 we apply this general formalism in cosmology. We argue that it
introduces a natural arrow of time. The relation between this and the
thermodynamic arrow of time must be derived at a later stage. We suggest
that a random, time dependent dynamics is the proper description of physics
near the Big Bang, and propose a particular class of random Hamiltonians for
this purpose, with results outlined above.

In the conclusions we recall the outline of our heuristic description of
holographic cosmology and its application to observational cosmology. We
sketch a program for deriving the assumptions and parameters of the
heuristic picture from the mathematical formalism presented in this paper.
We also introduce a more general model which describes a \textquotedblleft
gas of causally disconnected, asymptotically de Sitter (dS) universes"
embedded in a $p=\rho $ background. Such a model can implement the anthropic
principle for the cosmological constant, without requiring other parameters
of low energy physics to be anthropically selected.

\section{Local framework for a holographic theory of quantum gravity}

Thirty years of work on perturbative and non-perturbative formulations of
string theory, have presented us with ample evidence for the holographic
nature of this theory of quantum gravity. Every gauge invariant quantity in
all versions of the theory, refers to an observable associated with the
conformal boundary of a spatially infinite space-time.

There is a simple intuitive argument, which suggests why this should be the
case. A theory of gravitation must describe the apparatus which might
measure any given prediction of the theory, because all physical objects
gravitate. In a quantum theory, this is problematic, because the
mathematical predictions of quantum theory refer to limits of measurements
made by an arbitrarily large measuring apparatus. In a theory of gravity,
such a measuring apparatus would have large effects on the system being
measured unless it were moved an infinite distance away. This suggests that
the pattern we have observed in string theory is an inevitable consequence
of the marriage of gravitation and quantum mechanics. \textit{All gauge
invariant observables in a quantum theory of gravity describe the response
to measurements made by infinite machines on infinitely distant surfaces.}
String theory in asymptotically flat, asymptotically AdS, and asymptotically
linear dilaton space-times obeys this rule.

Stringy evidence and simple physical intuition thus both point to the
impossibility of defining gauge invariant quantities for local systems. But
the necessity of describing a real world, which is cosmological in nature,
suggests that we need a more local description of physics. This can be
reconciled with the arguments above only by recognizing that no local
description will be gauge invariant.

Indeed, this is a lesson we have already learned from attempts to quantize
gravity in the semi-classical approximation. In order to define a concept of
time and a quantum mechanics with unitary time evolution in this framework,
we must choose a classical background solution \cite{rubetal}\footnote{%
The examples of the relativistic particle and world sheet string theory
(viewed as two dimensional gravity) show that one can quantize a generally
covariant system beyond the semiclassical expansion only by second
quantizing it. This evidence suggested the notion of Third Quantization, but
there is no consistent formulation of a Third Quantized theory above two
dimensions. Practitioners of loop quantum gravity have also encountered the
unitarity problem of the Wheeler DeWitt equation. They tend to either put it
off to future research, or try to live with non-unitary time evolution.}.
The background plays the role of the infinite measuring device that we need
to define a gauge invariant notion of time. The resulting formalism is
quantum field theory in curved space-time. Time evolutions defined by
different classical solutions, or even by different coordinatizations of the
same classical solution, do not commute with each other and cannot be easily
reconciled. This leads to the notion of Black Hole Complementarity, which
gives a conceptual (though not yet a mathematical) resolution of the black
hole information paradox. Two of us (TB and WF) generalized this to
Cosmological Complementarity for Asymptotically dS (AsdS) space-times, and
E. Verlinde has suggested the name Observer Complementarity to describe
general space-times with event horizons.

Quantum field theory in curved space-time leads to the familiar paradox of
black hole decay, and fails decisively in the presence of space-time
singularities. The evidence is that the same is true for weakly coupled
string theory, which also relies on a classical space-time background. We
need a better way.

For some time, the present authors have felt that the fundamental clue to a
local formulation of quantum gravity could be found in Bousso's general
formulation of the holographic principle \cite{raph}. A fundamental notion
in Lorentzian geometry is the concept of causal diamond. This is the region
of intersection of the causal past of a point $P$ with the causal future of
a point $Q$ which is in the causal past of $P$. The covariant entropy bound
implies that for any causal diamond, the entropy that can flow through its
boundary is bounded by the area of the maximal area $d-2$ surface on the
boundary. We have conjectured \cite{tbmillmcosmoholcosm} that in the quantum
theory of gravity, this entropy should be associated with the logarithm of
the dimension of the Hilbert space necessary to describe all measurements
done inside the causal diamond. In every Lorentzian space-time, the
covariant entropy bound for a causal diamond, is finite for sufficiently
small time-like separation between $P$ and $Q$.

Of course, finiteness of the entropy of a density matrix does not by itself
imply that the Hilbert space of the system is finite. But finite entropy
density matrices in infinite systems, rely on special sets of operators
(typically the Hamiltonian) whose spectrum defines a natural restriction of
the Hilbert space. Our general discussion of quantum gravity suggests that a
local description should contain no such special operators. That is, in
general we expect the Hamiltonian of a local observer to be time dependent,
and different observers will have different, generally non-commuting, time
dependent Hamiltonians. The only natural density matrix, whose definition
does not depend on a special operator, is the unit matrix.

The finite dimensional Hilbert space conjecture meshes with the arguments
above, because a finite dimensional system cannot describe the infinite
machines which make operational sense of the precise mathematical
predictions of quantum theory. Thus we view a small causal diamond as
defined in quantum theory by a (generally time dependent) Hamiltonian on a
finite dimensional Hilbert space. Since such a system can never make
arbitrarily precise measurements on itself, its Hamiltonian and other
observables cannot be fixed. That is, given the \textit{a priori }%
restriction on the precision of measurements, we will always be able to find
many alternative mathematical descriptions, which agree up to the specified
level of precision allowed by the size of the causal diamond. We view this
statement as the quantum origin of the \textit{Problem of Time} in
semiclassical general relativity\footnote{%
More generally, it is the quantum origin of general coordinate invariance.}
and we view any given Hamiltonian description of a causal diamond as a gauge
fixing. The aptness of this metaphor will become more apparent as we get
deeper into the formalism.

We have not yet pointed out the most important aspect of our conjecture,
namely that it provides a derivation of a notion of locality from the
holographic principle itself. Indeed, what could it mean to assert the
finiteness of the operator algebra associated with a causal diamond, if not
the statement that it formed a tensor factor of the operator algebra of the
entire space-time? The operators of the causal diamond $\mathcal{D}$ commute
with all other operators necessary to describe the physics in any larger
causal diamond $\mathcal{D}^{\prime}$ containing $\mathcal{D}$.

The algebraic formulation of quantum field theory similarly assigns an
operator algebra to each causal diamond. The field theory operator algebras
are all infinite, and the detailed relation between algebraic and space-time
structure will be different than what we propose here. However, the
similarities of the two frameworks may eventually provide us with a better
understanding of how field theory arises as a limit of a real theory of
quantum gravity.

\subsection{The hilbert space of an observer}

The basic idea of our program is to use the holographic conjecture about the
dimension of the Hilbert space of a causal diamond, to translate geometrical
concepts into quantum mechanics. We urge the reader to think of the
geometrical pictures as ``guides to the eye" at this stage, and to think of
the quantum formalism as fundamental. At a later stage, one would hope to
obtain a mathematical derivation of the rules of Einsteinian geometry from
the quantum formalism. In this paper, we will provide one example of such a
derivation, in a very special case.

We will use the word \textit{observer} to denote a large, localized quantum
system, which is capable of carrying out ``almost classical" measurements on
its environment. Any such observer will follow a timelike trajectory through
space-time. We can describe this trajectory in terms of causal diamonds in
the following manner. First consider space-times such that the observer's
trajectory has infinite timelike extent in both past and future. Pick a
point $P$ on the trajectory and a segment of equal length to the past and
future of $P$. Take the causal diamond defined by the endpoints of this
segment. As we make the interval smaller, the FSB area of this diamond gets
smaller. If we want to associate this area with the logarithm of the
dimension of a Hilbert space, this process must stop at some smallest
length. Let $\mathcal{K}$ be the dimension of this smallest Hilbert space.
We will make a proposal for $\mathcal{K}$ in a moment.

Now we extend the interval around the point $P$, until the area of the
causal diamond has increased by the logarithm of the dimension of $\mathcal{K%
}$\footnote{%
One could imagine a formalism in which one changes the dimension of the
Hilbert space by one at each step. It is harder to describe this in terms of
an attractive operator algebra. Our motivation for tensoring in a fixed
Hilbert space at each step is the concept of a holographical pixel, to be
defined below.}. By continuing this procedure, we describe the information
that can be measured in experiments done by an observer in terms of a
sequence of Hilbert spaces, $\mathcal{H}_{N}$ of dimension $(\mathrm{dim}%
\mathcal{K})^{N}$. This corresponds to a sequence of causal diamonds, as
shown in Figure 1. The entropy of the maximally uncertain density matrix for
this system is $N\ln (\dim \mathcal{K})$. This is to be identified with one
quarter of the area of the causal diamond in Planck units.

For Big Bang cosmologies, we can do something similar, but it is convenient
to choose causal diamonds whose past tip lies on the Big Bang, and extend
them only into the future. The smallest causal diamond for any observer, is
that observer's view of the Big Bang hypersurface. Note that it will be
completely finite. In our view, the Big Bang looks singular in general
relativity, because one is thinking of the theory as a field theory and
trying to describe all of the degrees of freedom of that theory in each
horizon volume. The holographic principle suggests instead that near the Big
Bang surface, small causal diamonds contain very few degrees of freedom, and
have a completely non-singular quantum description.

Although the quantum mechanics of a causal diamond is always independent of
that in other causal diamonds in the same space-time\footnote{%
That is, the Hilbert space of a causal diamond contains all the degrees of
freedom necessary to describe measurements in that region. There \textit{will%
} be mappings between the Hilbert spaces of different causal diamonds, and
consistency relations among the different time evolution operators.}, one
should not imagine that the initial state in a generic causal diamond is
pure. Interactions to the past of the diamond could have entangled its
degrees of freedom with those of other disjoint diamonds. Our fundamental
cosmological hypothesis will be that the state in a causal diamond whose
past tip is on the Big Bang, is pure. This corresponds to the familiar
notion of \textit{particle horizon}. All quantum correlations between the
degrees of freedom of the system are to be generated by the dynamics, rather
than put in as initial conditions.

We would like to emphasize that this hypothesis introduces the Arrow of Time
as a fundamental input to the definition of cosmology. That is, we could
define both Big Bang and Big Crunch cosmologies (with, for simplicity, a
past or future with the asymptotic causal structure of Minkowski space), in
terms of semi-infinite sequences of Hilbert spaces. However, in the Big Bang
case, the initial conditions would be subject to our purity constraint for
causal diamonds whose tip lies on the singularity. By contrast, in the Big
Crunch, the initial conditions would be described in terms of scattering
data in the remote past. Even if we discussed finite causal diamonds whose
future tip lay on the Big Crunch, it would not make sense to assume the
final state in those causal diamonds was pure. It has been correlated with
the states in each other causal diamond, by the evolution of the scattering
data down to the singularity. \textit{Thus we contend that the intrinsic
formulation of a theory of quantum cosmology, forces us to introduce a time
asymmetry, when there is a cosmological singularity}\footnote{%
If there is a reasonable description of a universe which undergoes a Big
Bang followed by a Big Crunch, the time direction will be specified by the
purity constraint. We would describe such a universe in terms of pure states
in causal diamonds with their tip on the Big Bang. The range of $N$ would be
finite, and only the last causal diamond in the sequence would touch the Big
Crunch.}.

The causal diamond formalism automatically introduces an ultraviolet energy
cutoff, because it discretizes the time step. Notice however that the cutoff
is not uniform in time. In a region of space-time (and a given foliation)
where the spatial curvature is negligible, the area of causal diamonds
scales like the proper time to the $d-2$ power. So a fixed area cutoff,
corresponds to a finer and finer slicing of proper time, as $N$ increases.
To get an intuitive feeling for this scaling note that it is the same as
what one gets by applying the time energy uncertainty relation and saying
that the time step is the inverse of the energy of the largest black hole
that can fit in to the causal diamond at step $N$\footnote{%
Here and henceforth, we will use a rough definition of a black hole as a
localized concentration of energy and entropy, which maximizes the entropy
for a given energy. We are aware that none of these concepts has an
absolutely rigorous definition in general relativity.}.

Note that, while we have introduced geometrical notions (area), our
construction says nothing as yet about the actual geometry of space-time.
One can introduce trajectories via sequences of causal diamonds with fixed
area step, in any Lorentzian space-time. Certain global aspects of the
space-time are encoded in the behavior of $\mathcal{H}_N$ for large $N$. In
space-times with asymptotic causal structure like that of Minkowski space,
the area of the causal diamond goes to infinity continuously as the
time-like separation between its tips goes to infinity. In asymptotically
AdS space-times, the area goes to infinity at finite time-like separation,
when the causal diamond hits the time-like boundary of AdS. After that point
the operator algebra becomes infinite and is equal to the algebra of
conformal fields on the boundary, smeared with functions of compact support
in boundary time. In asymptotically dS spaces, we expect the operator
algebra to remain finite even in the limit of infinite proper time. We have
already discussed the modification of the formalism necessary to the
description of space-times with cosmological singularities. Thus, the
boundary geometry of space-time affects the nature of the index set $N$ (in
AdS, the mapping between $N$, which counts area, and time, becomes singular
at a finite time. After this point, the time becomes a continuous parameter
while the area is infinite). In asymptotically dS space-time we can choose $%
N $ to parametrize a discrete global time. Then $N$ is allowed to go to
infinity, but we stop adding degrees of freedom at a finite value of $N$).
More generally, we expect the geometry to emerge from an interplay between
area and the time evolution operators in each Hilbert space $\mathcal{H}_N$.

In each Hilbert space, we postulate a sequence of unitary operators
\linebreak $U_N (k) \equiv e^{-i H_N (k)}$ for $1 \leq k \leq N$. In a Big
Bang space-time $U_N (k)$ is supposed to represent the evolution of the
system between the future tips of the $k$-th and $(k-1)$-th causal diamond.%
\footnote{%
From now on we will concentrate on the cosmological case. Much of the
discussion has an obvious generalization to other boundary conditions.}
Here we encounter the first of the fundamental consistency conditions of
quantum gravity. The Hilbert space $\mathcal{H}_N$ contains a tensor
factor isomorphic to $\mathcal{H}_K$ for $K < N$. Inside this factor the
dynamical description of the later observer, must coincide with its own
past history. That is
\begin{equation}
U_N(k) = U_K(k) \otimes V_{NK}(k),  \label{consista}
\end{equation}
for $k \leq K$. We should view the operator $V_{NK}(k)$ as describing the
dynamics of degrees of freedom, which are, at time $k$, not observable by
the observer under discussion. It acts only on the tensor complement of $%
\mathcal{H}_K$ in $\mathcal{H}_N$. It will become important when trying to
make the dynamics consistent with the descriptions given by other observers.

We hope that this discussion of the Hilbert space of a single observer has
been relatively easy to follow. By contrast, it is extraordinarily difficult
to get one's head around the consistency conditions relating observers with
different time-like trajectories. We attack this question by first
introducing the $p=\rho$ cosmology, where there is a simple solution of all
of the consistency conditions. Only at the end of our discussion of this
cosmology will we return to the consistency conditions in a general
space-time. First however, we introduce our parametrization of the operator
algebras in terms of holographic pixels, and define the Hilbert space $%
\mathcal{K}$.

\subsection{SUSY and the holoscreens: the degrees of freedom of quantum
gravity}

We now want to make an ansatz for the Hilbert space $\mathcal{K}$ which will
connect our formalism to Riemannian geometry. If we associate the degrees of
freedom with the holographic screen of a causal diamond, then the most
fundamental thing that occurs when we increase the size of the diamond is
that we ``add a pixel" to the screen. The minimal new information must tell
us about the size and orientation of that pixel, and about the null
direction along which information from the bulk is projected onto the pixel.

There is a classical geometrical description of the orientation of a
holographic screen in terms of \textit{pure spinors }\cite{cartpen}. A pure
spinor in $d$ dimensions satisfies
\begin{equation}
\bar{\psi}\gamma ^{\mu }\psi \gamma _{\mu }\psi =0  \label{pure}
\end{equation}%
The defining equation is homogeneous and classically one views two pure
spinors as identical if $\psi _{1}=\lambda \psi _{2}$, where $\lambda $ is
real or complex depending on the reality of the spinor representation. In $%
3,4,6$ and $10$ dimensions, a general spinor in the smallest irreducible
spinor representation of the Lorentz group is automatically pure. The CP
equation comes up repeatedly in superstring theory, particularly in the
\textit{super-embedding} approach \cite{sorokin}.

The CP equation defines neither the position nor the size of the holographic
pixel. Only the direction of the null vector and the orientation of its
screen are fixed. This is in accord with the intuition that metrical
notions, like area, are measured in Planck units, and should not appear
until we quantize the theory.

To quantize the pixel variable $\psi$, we first note that it has half the
components of a general Dirac/Majorana spinor (we impose Majorana conditions
in those dimensions in which they exist). Denote the non-vanishing
components as $\hat{S}_a$. They transform as the spinor representation of $%
SO(d-2)$, the transverse rotation group which leaves $n^{\mu} $ invariant.
Note that in choosing to quantize only the physical components of the pure
spinor, we are partially choosing the gauge for local Lorentz invariance,
leaving over only an $SO(d-2)$ subgroup. Quantization of the pixel variable
is dimension dependent. In the remainder of the paper, we will treat $p=\rho$
universes with arbitrary dimension, but in order to be specific, we will
here discuss only the case $d=11$ , where $\hat{S}_a$ has $16$ real
components. The finite Hilbert space $\mathcal{K}$ of the previous section
will be identified with the Hilbert space of a single quantized pixel. The
most general $SO(9)$ invariant quantization rule, which is representable in
a Hilbert space with a finite number of states is

\begin{equation}
\lbrack \hat{S}_{a},\hat{S}_{b}]_{+}=2\delta _{ab}  \label{quant}
\end{equation}

Note that this rule breaks the projective invariance of the classical CP
equation, except for a $Z_2$ subgroup. We view this residual $Z_2$ as a
gauge symmetry, which should be implemented in the quantum theory.

We now utilize these variables to construct the Hilbert spaces of the
previous section. For a single observer we add a single copy of the $\hat{S}%
_a$ algebra at each time step. The new operators, $\hat{S}_a (N)$, commute
with the operators, $\hat{S}_a (t) ; t < N$, describing the smaller causal
diamond at the previous time step. The Hilbert space we tensor in is the
irreducible representation of this Clifford algebra. It is easy to satisfy
the
consistency conditions for the evolution operators, by choosing $H_N (k)$, $%
N > k$, to be a sum of two terms. The first depends only on the $\hat{S}_a
(t)$ for $t \leq k$, and the second only on those with $t > k$. The first
term is chosen equal to $H_k (k)$.

$Z_2$ gauge invariance is guaranteed by choosing each Hamiltonian to contain
only even polynomials in the pixel operators. We can then perform a $Z_2$
gauge transformation, to define new variables by

\begin{equation}
S_a (n) = (-1)^{F_n} \hat{S}_a (n),  \label{fermi}
\end{equation}
where $(-1)^{F_n}$ is the product of all of the $\hat{S}_k$ for $k \neq n$.
We then obtain the fermionic algebra

\begin{equation}
\lbrack {S}_{a}(m),S_{b}(n)]_{+}=2\delta _{ab}\delta _{mn}
\label{fermiquant}
\end{equation}

Fermi statistics is thus seen to be a quantum remnant of the projective
invariance of the CP equation, and the spin statistics connection is built
in to our formalism\footnote{%
The cosmology we will describe in this paper has no particle excitations, so
the relation between these fermionic commutation relations and the
statistics of particles will not be evident. In \cite{tbholosugra} one of
the authors will present a holographic description of 11 dimensional SUGRA
in flat space-time which will exhibit the precise connection.}

Later, when we speak of maps between Hilbert spaces corresponding to
spatially separated, but overlapping causal diamonds, $\mathcal{H}(\mathcal{D%
}_{1})$ and $\mathcal{H}(\mathcal{D}_{2})$ we will view these maps as
implemented by isomomorphisms between subalgebras of the pixel operators on
each Hilbert space. Note that these need not be linear mappings between the
generators. We can find non-linear functions of the pixel operators, which
satisfy the same Clifford algebra. The homomorphism might be a linear map
between the fundamental pixel operators of one Hilbert space, and such
\textquotedblleft composite" pixel operators in another.

\subsection{Rotation invariance}

A model of a homogeneous isotropic universe, should be invariant under
spatial rotations. In our $11$\textit{D} example, the $16$ real $S_{a}$
operators transform as a spinor of $SO(9)$ but not of $SO(10)$. There is an
analogy, which we believe will be helpful in understanding rotation
invariance \cite{tbholosugra}, between the $S_{a}(n)$ operators and sections
of the spinor bundle over the 9-sphere. Any such section is given locally,
by a map $S_{a}(\mathbf{\Omega })$, from the sphere to the spinor
representation of the $SO(9)$ which preserves a point $\mathbf{\Omega }$. We
should think of the $S_{a}(n)$ as finite dimensional analogs of sections of
the spinor bundle over the sphere.

The seminal idea of non-commutative geometry \cite{connes} is to replace the
commutative $C^*$ algebra of continuous complex valued functions on a
manifold, with a general non-commutative $C^*$ algebra. In particular, if we
choose finite dimensional matrix algebras we obtain \textit{fuzzy spaces}.
Particular infinite sequences of matrix algebras lead to fuzzy
approximations to Riemannian manifolds.

In non-commutative geometry, the concept of vector bundle is replaced by the
(equivalent in the commutative case) notion of a \textit{projective module}.
A projective module $R$ over an associative algebra $\mathcal{A}$ is a
representation of $\mathcal{A}$ with the property that there exists another
representation $\bar{R}$ such that $R \oplus \bar{R} = \mathcal{A}^p$, where
the power means $p$th tensor product of the regular representation of $%
\mathcal{A}$ on itself by left multiplication. This is the analog of the
existence of an anti-bundle $\bar{V}$ for each vector bundle $V$ over a
commutative manifold, such that $V \oplus \bar{V}$ is trivial.

Our $S_a (n)$ variables should belong to an operator valued projective
module for a finite dimensional associative algebra on which $SO(10)$ acts.
Finite dimensional representations of the Clifford-Dirac algebra $\gamma^M$
of SO(10) are examples of such fuzzy $9$ spheres. The smallest one is given
by the irreducible representation of the Clifford-Dirac algebra and has real
dimension $32$. In formulas below, we will use this doubling of the indices
of $S_a (n)$ to ensure $SO(d-1)$ rotation invariance.

We will not pursue these rotational properties further in this paper, but
note merely that they may be helpful in resolving a puzzle we will encounter
later.

\section{Quantum cosmology of a dense black hole fluid}

\subsection{The random operator ansatz}

We now want to present a complete solution of the general constraints on
quantum cosmology. We will argue that this solution corresponds to a flat
FRW universe with equation of state $p=\rho$. This is the system which we
have studied heuristically in previous publications under the name of ``a
dense black hole fluid". The mathematical analysis of this section will, we
believe, amply justify that colorful terminology. We emphasize that we are
presenting this solution of the constraints before making a general
statement of what the constraints are. We hope that this order of
presentation will help readers to understand the general construction.

A fundamental clue to our mathematical formalism is the result of BKL \cite%
{BKL} that the dynamics of general relativity near a space-like singularity
is chaotic. This suggests that the quantum theory should be described by a
random Hamiltonian. The causal diamond formalism and its description in
terms of fermionic holopixels suggests a particular ensemble of random
Hamiltonians.

Let us begin by considering the quadratic term in the Hamiltonian $H_N (N)$.
It has the form

\begin{equation}
H_{N}^{(2)}(N)=i{\frac{1}{N}}S^{a}(n)h^{mn}S^{a}(m)\equiv {\frac{1}{N}}%
H_{FT},  \label{randham}
\end{equation}%
where $h^{mn}$ is a real anti-symmetric $N\times N$ matrix. We have imposed $%
SO(d-1)$ invariance by using the invariant scalar product on the component
indices of $S$\footnote{%
Here we are assuming that the appropriate fuzzy spinor bundle is just the
direct sum of copies of the minimal one in which we double the indices of $%
S_{a}$ to extend it to an $SO(d-1)$ representation. This doubling should be
understood in the above formula. It may be that this missing factor of $d-2$
we encounter below is an indication that this is the wrong choice.}. Our
ansatz will be to choose $h$ to be a gaussian random matrix with the
standard probability distribution $P(h)=e^{N\mathrm{tr}h^{2}}$. For large $N$
the distribution is described by a master field, with spectral density given
by the Wigner semi-circle law, $\rho _{h}(x)=\sqrt{1-x^{2}}$. The
distribution is flat near the origin and has a cutoff of order one for its
eigenvalues. It then follows that the large $N$ thermodynamics of $%
H_{FT}\equiv NH_{N}^{(2)}(N)$ is that of a free $1+1$ dimensional fermionic
field theory \cite{KapluWein}. The entropy is of order $N$, the eigenvalue
spacing is of order ${\frac{1}{N}}$. Thus $H_{FT}$ should be viewed as a $%
1+1 $ dimensional free fermion system with UV cutoff of order $1$, living on
an interval of length of order $N$. The $1+1$ dimensional entropy and energy
densities are related by $\sigma _{1+1}\propto \sqrt{\rho _{1+1}}$. We will
identify these as the space-time entropy and energy densities of our
cosmology. This equation of state would be appropriate for an FRW universe
with equation of state $p=\rho $. Before pursuing this relationship, let us
extend our ansatz for the basic Hamiltonian.

The thermodynamics of this system is dominated by the IR physics of $1+1$
CFT. This will be unchanged by a wide class of perturbations of $H_{FT}$.
Indeed, the only relevant perturbations of this system are the fermion mass
and the marginally relevant four fermi operators. Our random matrix ansatz
has automatically set the fermion mass to zero. The marginally relevant
perturbations are marginally irrelevant if their sign is appropriately
chosen. Thus we can add to $H_{FT}$ an arbitrary even function of the pixel
operators of degree $\geq 4$, \textit{whose coefficients in the eigenvalue
basis of $h_N$ are smooth functions of the eigenvalue in the large $N$ limit}%
, as long as the sign of the quartic terms is chosen correctly. We see that
a very wide class of random Hamiltonians for our system, will have identical
large $N$ thermodynamics. Thus, our full ansatz for the cosmological time
evolution is that for each $N$ we make an independent choice of random
Hamiltonian, $H_N (N)$, from the distribution defined in the last two
paragraphs.

The operators $H_N (k)$ with $k < N$ are partially fixed by the requirement
that $H_N (k) = H_k (k ) \otimes 1 + 1 \otimes O_N (k)$, where $O_N (k)$
depends only on the variables $S_a (t)$ with $N \geq t > k$. The universe
experienced by the observer in this causal patch is unaffected by the choice
of these operators. One might however have thought that they were
constrained by the spatial overlap conditions. For our choice of overlap
conditions in the $p=\rho$ universe, this turns out to be untrue. The $O_N
(k)$ are completely unconstrained. We suspect that this might not be the
case for more general space-times. We will see below, that although our
ansatz reproduces the scaling laws of the $p=\rho$ universe, it fails to
reproduce certain more refined features of the geometry. This leads us to
surmise that the ansatz needs to be modified. The necessary modification is
likely to require us to specify $O_N (k)$.

A full definition of a quantum space-time must include the descriptions of
other observers. A coordinate system can be thought of as a way of covering
space-time by the trajectories of observers. We will choose time-like
observers and will choose a time slicing such that at a given time, along
any trajectory defining our coordinate system, the area of the maximally
past extended causal diamond is the same. We call this equal area slicing of
a Big Bang space-time. At (say) the initial time the ends of the
trajectories form a lattice. We specify the topology of this spatial slice,
including its dimension by choosing a particular topological lattice. For
simplicity of exposition, we will choose the $d-1$ dimensional hypercubic
lattice. At large $N$ this choice will not matter and our ansatz would work
for any lattice with the same continuum topology.

Each trajectory is specified by a sequence of Hilbert spaces and unitary
operators as above. Two neighboring trajectories would correspond to two
overlapping sequences of causal diamonds, as shown in Fig. \ref%
{intersectionnesteddiamonds} . \textit{A priori} one could imagine making
independent choices of Hamiltonian at each point on the spatial lattice. We
will argue that this is inconsistent with the random operator hypothesis,
and that in fact the sequence of Hamiltonians defining a given observer will
be identical at all spatial points. Only the initial state can differ from
point to point. Indeed, the causal diamonds of two trajectories will
generally have an overlap Fig. \ref{intersectionnesteddiamonds} . The
overlap will not be a causal diamond, but will contain some maximal area
causal diamond. It is reasonable to postulate that the information which
could be accessed in the overlap can be encoded in a Hilbert space which is
(isomorphic to) a tensor factor in each of the individual causal diamond
Hilbert spaces. Furthermore, if we look at the actions of the time evolution
operators of the individual diamonds, on this common factor space, they must
agree. Since there are many such overlaps, this is a very strong constraint
on the dynamics.

In the $p=\rho $ cosmology, our ansatz for spatial overlap Hilbert spaces is
simple and general. If we consider two Hilbert spaces $\mathcal{H}_{N}(%
\mathbf{x)}$ and $\mathcal{H}_{N}(\mathbf{y)}$ which are $s$ steps away from
each other on the lattice, we choose the overlap to be $\mathcal{H}_{N-s}(%
\mathbf{x)=\mathcal{H}_{N-s}(y)}$. In finer detail, we identify the
individual $S_{a}(t,\mathbf{x})$ operators, with their counterparts in the
Hilbert space at $\mathbf{y}$. If we now require that the Hamiltonian
evolutions of each sequence of causal diamonds are identical, then all of
our consistency conditions are satisfied, in the following sense. For each
geometrical overlap between causal diamonds, we have defined a Hilbert space
and a sequence of time evolution operators, which purports to describe the
physics in the overlap region of space-time. The overlap Hilbert space is a
tensor factor in each of the individual observer's Hilbert space.
Furthermore, the dynamics in this tensor factor is consistent with that
defined by either of the individual observers.

It seems likely, but we have not been able to prove, that there is no other
solution of the overlap conditions which would be compatible with each
observer having a random sequence of Hamiltonians.

\subsection{Homogeneity, isotropy and flatness}

Our construction is homogeneous on the spatial lattice. We have built
isotropy into our construction in a formal way, by insisting on $SO(10)$
invariance. The overlap rules give us further indications that our system is
isotropic. We will have occasion to refer both to the Euclidean distances
and angles on our hypercubic lattice, and the actual Riemannian distance in
the space-time metric we claim to be constructing. The reader should be
careful to keep these two ideas completely separate. We have defined a
space-time lattice with lattice points labeled $(N,\mathbf{x})$. Define the
base of the causal past of the point $(N,\mathbf{x})$ to be the set of all
points on the lattice, whose Hilbert space at time $N$ has an overlap with $%
\mathcal{H}_{N}(\mathbf{x})$. According to our overlap rules, the boundary
of this set is given by the endpoints of walks on the lattice, starting at $%
\mathbf{x}$ and increasing the Euclidean distance on the lattice at each
step. The base of the causal past thus forms a hypercube oriented at forty
five degrees to the coordinate axes. Each step along the walk reduces the
area of overlap by one unit, and so should be thought of as increasing the
Riemannian distance by some ($N$ dependent) unit. Thus, the boundary of the
base of the causal past consists of points which are the same Riemannian
distance away from \nolinebreak $\mathbf{x}$.

Think of a \textit{carpenter's ruler} which follows a walk along the lattice
to the boundary of the base of the causal past. The map between the
coordinate (lattice) space and the real geometry, is given by
``straightening out the carpenter's ruler". The tilted hypercube is mapped
into a sphere.

We have thus derived homogeneity and isotropy of our cosmology from our
definition of the overlap rules. Given the non-compact topology of the
lattice, the spatial curvature is non-positive. There are three different
arguments that it is zero. The first is simply that our model saturates the
entropy bound. At any given late time, the excited states of our system are
generic states of the Hilbert space, because they are obtained by the action
of a sequence of random Hamiltonians. We know that even the maximally stiff
equation of state $p=\rho$ cannot saturate the entropy bound in a universe
of negative curvature.

The second argument for flatness also shows us that our spin connection is
Riemannian. The overlap conditions have forced us to identify the $S_a$
operators in Hilbert spaces at different points. Thus, the parallel
transporter is the identity in $SO(10)$ and the curvature of the spin
connection vanishes.

Finally, note that for large $N$ the spectrum of our system has a scaling
symmetry because it is that of a $1+1$ CFT. If it is to be identified with
an FRW universe, that universe should have a conformal isometry
corresponding to the symmetry\footnote{%
We also learn that the ``matter" in this universe must be invariant under
this conformal isometry.}. Such an isometry exists for any FRW universe with
flat spatial sections and a single component equation of state. Curved
spatial sections introduce a scale and such geometries do not have a
conformal isometry.

The last argument can be stated in another way. We have defined a sequence
of physical spheres, \textit{the causal boundaries at time $N$} on our $d-1$
dimensional coordinate lattice. If the spatial geometry were curved, we
would expect to see a scale, the radius of curvature, at which the behavior
of the geometry changed. As we take $N$ to infinity we will sweep through
this scale. However, the dynamics does not have such a scale in it. It
becomes scale invariant for large $N$.

To summarize, we have shown that the random Hamiltonian ansatz, which obeys
our consistency conditions for a quantum cosmology, gives a spatial geometry
which is homogeneous, isotropic and flat. It also obeys two laws which
suggest that it is in fact the quantum realization of $p=\rho$ cosmology.
The entropy bounds are saturated for all time, and the energy entropy
relation of an extensive $p=\rho$ fluid is valid at all times. In the next
subsection we will provide further evidence that this is the right
interpretation of our system.

\subsection{Time dependence - scaling laws}

In order to discuss the time dependence of our geometry, we have to identify
the conventional cosmological time parameter in terms of the parameters of
our quantum system. In any flat FRW cosmology, the area of causal diamonds
at cosmological time $t$, scales as $t^{d-2}$. Thus, we should write $N\sim
t^{d-2}$. The logarithm of the $N$ dependent time evolution operator is $%
-i\Delta NH_{N}$, where $\Delta N$ is $N$ independent. Writing
\begin{equation}
\Delta N\sim t^{d-3}\Delta t  \label{transf}
\end{equation}%
we see that the cosmological time dependent Hamiltonian is
\begin{equation}
H(t)\sim N^{\frac{{(d-3)}}{{(d-2)}}}H_{N}  \label{cosmtime}
\end{equation}
$H_{N}(N)$ is the Hamiltonian as viewed by an observer in a given causal
diamond. To the extent that one can really talk about such an observer in
the heuristic picture of a dense black hole fluid one views it as hovering
about the maximal black hole at a distance of order its Schwarzchild
radius. The energy of the system is just the energy of the black hole for
such an observer. In our quantum mechanical model, for most states of that
system, the energy per unit length is of order $1$ (\textit{i.e.} $N$
independent). Thus
\begin{equation}
H(t)\sim N^{\frac{{(d-3)}}{{(d-2)}}}  \label{horbh}
\end{equation}%
This implies that the local cosmological observer sees an energy which
scales like the mass of the maximal black hole, exactly as required by our
heuristic picture. Note that this calculation works in any dimension.

We can get further confirmation by noting that we have outlined an order of
magnitude calculation of the physical size of the particle horizon in the
previous subsection. It is $N$ lattice steps in coordinate space, while the
UV cutoff scales like $N^{-{\frac{{(d-3)}}{{(d-2)}}}}$. Thus, the physical
size of the particle horizon scales like $N^{-{\frac{1}{{d-2}}}}$. Since the
spatial geometry is flat, this implies a horizon volume
\begin{equation*}
V_{H}\sim N^{-{\frac{{d-1}}{{d-2}}}}
\end{equation*}%
The cosmological energy density is obtained by dividing \ref{horbh} by this
volume. Thus,

\begin{equation}
\rho \sim N^{-{\frac{2}{{d-2}}}}\sim {\frac{1}{t^{2}}}  \label{rho}
\end{equation}%
where at the last stage we have again used the relation between entropy and
cosmological time. Similarly, the total entropy is $N$ so the entropy
density is

\begin{equation}
\sigma \sim N^{-{\frac{1}{{d-2}}}}\sim {\frac{1}{t}}  \label{sigma}
\end{equation}%
Thus, we have obtained both the $\sigma \sim \sqrt{\rho }$ equation of state
of the $p=\rho $ universe, as well as the ${\frac{1}{t^{2}}}$ dependence of
energy density, usually derived from the Friedmann equation, from a purely
quantum mechanical calculation.

\subsection{Time dependence: a consistency relation, and a failure}

Another interesting geometrical quantity is the area of the overlap causal
diamond, as a function of $N$ and of the geodesic separation between the
trajectories. In the Appendix \ref{Appendix A} we calculate this area for a
general flat FRW space-time. Not surprisingly, it scales like $\Delta ^{d-2}$
where $\Delta $ is the geodesic separation. On the other hand, in our
quantum definition of overlap, the entropy in the overlap is $(N-k)L_{S}$,
where $k$ is the minimal number of lattice steps separating the tips of the
two causal diamonds, and $L_{S}=\ln (\dim \mathcal{K})$. The overlap entropy
is linear in $k$. We have argued that for fixed $N$, the number of steps is
linear in the geodesic distance $\Delta $.

This is not necessarily a contradiction. The quantum calculation is only
supposed to agree with the geometrical picture in the limit that $N$ is
large, and for causal diamonds which have large area. The area of the
overlap diamond decreases to zero as $k \rightarrow N$. Thus, it might be
reasonable to require agreement with geometry only for ${\frac{k}{ N}} \ll 1$%
. In this limit, both expressions are linear in $k$ and we can compare how
they scale with $N$.

Consider two diamonds in a flat FRW space-time, whose future tips lie at
conformal time $\eta _{0}$. Let these two diamonds be separated by co-moving
coordinate distance $\Delta x$. Then, according to our calculations in the
Appendix \ref{Appendix A}, the area of the maximal causal diamond which fits
in their intersection is, to leading order in $\Delta x$,

\begin{equation}
A_{int}^{Geo}=A\left( 1-{\frac{{d-2}}{\eta _{0}}}\Delta x\right)
\label{geomint}
\end{equation}%
To fit with the quantum mechanical picture, where the entropy associated
with this intersection is $(N-k)L_{s}$ for two diamonds separated by $k$
lattice steps , we must choose $A=4NL_{s}$, and ${\frac{N}{\eta _{0}}}%
(d-2)\Delta x=1$, for the co-moving separation corresponding to a single
step on our coordinate lattice. The geodesic distance at time ${\frac{\eta
_{0}}{2}}$ (the time of maximal area on the causal diamonds) represented by
a step is thus
\begin{equation}
\Delta d=a \left( \frac{\eta _{0}}{2} \right) \frac{\eta _{0}}{N(d-2)}
\label{dstep}
\end{equation}

There is now a consistency condition. We can compute the area of the causal
diamond at time $\eta_0$ in two ways. On the one hand, in order to causally
separate two causal diamonds, we must, according to our overlap rules, move $%
N$ steps on the lattice. This indicates that the radius of the maximal
sphere on the causal diamond is ${\frac{N}{2}}$ lattice steps. This
corresponds to an area

\begin{equation}
A=\Omega _{d-2}\left({\frac{N}{2}}\Delta d\right)^{d-2}  \label{ageom}
\end{equation}%
$\Omega _{d-2}$ is the area of a unit $d-2$ sphere. This area (in Planck
units, and we have set $G_{N}=1$) must be $4NL_{s}$. This gives us a second
equation for $\Delta d$

\begin{equation}
\Delta d={\frac{2}{N}\left(\frac{4NL_{s}}{\Omega _{d-2}}\right)}^{\frac{1}{%
d-2}}  \label{dstepb}
\end{equation}

Note that this has an attractive scaling property $\Delta d \sim N^{- {\frac{%
{d-3}}{{d-2}}}} $. We have suggested that the proper time cutoff scales like
the inverse of the energy of the maximal black hole, which fits in a causal
diamond. Here we find a spatial distance cutoff of the same order of
magnitude.

To compare the two expressions for $\Delta d$ we use the Friedmann equation
for $p=\rho$ geometry to write

\begin{equation}
a\left(\frac{\eta _{0}}{2}\right)=a_{0}\left(\frac{{(d-2)a_{0}\eta _{0}}}{{%
2(d-1)}}\right)^{{\frac{1}{{d-2}}}}  \label{amax}
\end{equation}%
We also express $\eta _{0}$ in terms of the area, and thence the entropy

\begin{equation}
\eta _{0}a_{0}=2^{\frac{{d+1}}{{d-1}}}\left(\frac{{NL_{s}(d-1)}}{{%
(d-2)\Omega _{d-2}}}\right)^{\frac{1}{{d-1}}}  \label{conftime}
\end{equation}%
Plugging these expressions into \ref{dstep} we obtain

\begin{equation}
\Delta d={\frac{1}{{d-2}}}{\frac{2}{N}\left(\frac{4NL_{s}}{\Omega _{d-2}}%
\right)}^{\frac{1}{d-2}}  \label{dstepc}
\end{equation}

Thus, the two expressions for the geodesic distance scale the same, but
differ by a factor $d - 2$. We have not been able to explain this
discrepancy. It is clearly related to the fact that the relation of overlap
area to geodesic distance in geometry is $A \sim \Delta^{d-2}$. We suspect
the discrepancy indicates the need for a slight modification of our overlap
rules, and is connected to another disturbing feature of these calculations.
One might have expected the numerical factors in the matching of geometry to
quantum mechanics would depend on the dimension of the pixel Hilbert space $%
\mathcal{K}$, which in turn depends on the space-time dimension. Further,
one might have expected the overlap rules to have a directional dependence
on the lattice which should break the local $SO(d-1)$ invariance of the
individual fermionic Hilbert spaces, leaving only a global $SO(d-1)$.
Neither of these expectations is realized in our current rules, and we
expect that when the rules are modified to take this into account, the
discrepant factor of $d - 2$ will disappear.

We emphasize that the calculation of the area of overlaps \textit{does} have
consistent scaling behavior with $N$. This is an independent check that our
quantum system satisfies the scaling laws of $p=\rho$ geometry. In order to
achieve this we had to insist on comparing geometric and quantum predictions
only at leading order in the area. For a more normal space-time background
this would probably not be sufficient to reproduce what we know of the
physics. The $p=\rho$ fluid appears to be a system in which the laws of
geometry are satisfied only in a very coarse grained sense.

We have tried to find other detailed numerical comparisons between our
quantum formalism and space-time physics. Unfortunately they all seem to
lead simply to a definition of constants in the quantum formalism. We record
these calculations in the Appendix \ref{Appendix A}.

\section{More general space-times}

The general kinematic framework for discussing holographic space-times is
very similar to what we outlined above. We will distinguish two different
kinds of temporal asymptotics: Scattering universes and Big Bang universes.
Big Crunch space-times pose additional problems, which we will ignore in
this paper.

A Scattering universe has past and future asymptotics which are describable
in terms of QFT in curved space-time. That is to say, in both the past and
the future there is a complete set of scattering states, which may be viewed
as localized excitations propagating on a classical geometry. The Penrose
diagram of a true scattering universe will be like that of Minkowski space,
or the universal cover of $AdS$. In the semi-classical approximation, dS
space is a scattering universe, but if one accepts the conjecture that the
quantum theory has a finite number of states, this is no longer precisely
correct. Nonetheless, we will include dS space under the rubric of
scattering universes. The reason for this is our belief \cite{mcosmo} that
as the c.c. goes to zero, the theory of dS space will contain a unitary
operator which converges to the scattering matrix of an asymptotically flat
space-time. The definition of this operator will contain ambiguities which
go to zero exponentially with the c.c., as long as the scattering energies
are kept fixed as $\Lambda $ goes to zero. We will reserve the phrase
\textit{true scattering universes} to describe space-times with a Penrose
diagram similar to that of Minkowski space. This does not imply that the
geometry is asymptotically flat. Non-accelerating FRW universes are also
true scattering universes. Big Bang space-times can asymptote either to a
future scattering universe or to dS space.

In a scattering universe, one describes the quantum theory by picking a
point on a time-like trajectory, and considering the causal diamonds defined
by successively larger intervals around that point, as in Fig. \ref%
{nesteddiamonds}. For each causal diamond we have a sequence of unitary
transformations $U_{N}(k)$ which describe time evolution in each of the
sub-diamonds contained in it. These must satisfy the causality requirement

\begin{equation*}
U_N (k) = U_k (k) \otimes W_N (k) ,
\end{equation*}
where $W_N (k)$ acts only on the tensor complement of $\mathcal{H}_k$ in $%
\mathcal{H}_N$. As $N \rightarrow \infty$ , in a true scattering universe,
we will have

\begin{equation*}
U_{N}(N)\rightarrow U_{+}(N)SU_{-}(N),
\end{equation*}%
where $U_{\pm }(N)$ describe free asymptotic propagation and $S$ is the
scattering matrix. In an asymptotically (past and future) dS universe there
should be, in the limit of small cosmological constant, a similar
construction \cite{mcosmo}\cite{nightmare}. However, in this case we cannot
take the large entropy limit. After some time, the dimension of the Hilbert
space stops increasing. Nonetheless, in the limit of small cosmological
constant, we expect an approximate S-matrix to exist. It would describe a
single observer's experience of excitations coming in through its past
cosmological horizon and passing out through its future cosmological
horizon. However, most of the states in the system cannot be viewed in this
way. From the point of view of any given observer, they are instead quantum
fluctuations bound to the cosmological horizon. The interaction between the
horizon states and the \textquotedblleft scattering states" introduces a
thermal uncertainty in the scattering matrix. This uncertainty cannot be
removed by local measurements, because the locus of the horizon states is an
extreme environment from the point of view of a given observer. It cannot
perform observations near the horizon without a large expense of energy,
which distorts the measurement \cite{bhcomplementarity}.

Thus, in the $AsdS$ case, the S-matrix is only approximately defined. Paban,
and two of the present authors \cite{nightmare} have argued that the
S-matrix for energies\footnote{%
In this sentence, energy refers to the eigenvalue of an operator which
approaches a timelike component of the momentum in the Poincare algebra, as
the c.c. goes to zero. This is not the same as the Hamiltonian of the static
observer, though the commutator between these generators is expected to be
small in the subspace with fixed Poincare energy.} that are kept fixed as
the c.c. goes to zero, should have a well defined but non-summable small $%
\Lambda $ asymptotic expansion, with errors of order (in four dimensions) $%
e^{-({\frac{M_{P}4}{\Lambda }})^{3/2}}$.

In both a true scattering universe, and an AsdS universe the description of
a single observer suffices from an operational point of view. However, the
constraints on the quantum mechanics of a single observer are not very
strong. As in the $p=\rho $ universe, we introduce other observers as a
lattice of sequences of Hilbert spaces $\mathcal{H}_{N}(\mathbf{x)}$. The
lattice has the topology of $R^{d-1}$ \footnote{%
Compact or partially compact spatial topologies present new difficulties,
with which we are not yet prepared to deal.}. For each pair of points on the
lattice, we introduce, at each $N$, a tensor factor $\mathcal{O}_{N}(\mathbf{%
x,y})$ of both $\mathcal{H}_{N}(\mathbf{x})$ and $\mathcal{H}_{N}(\mathbf{y}%
) $. For nearest neighbor points, the dimension of $\mathcal{O}_{N}(\mathbf{%
x,y})$, ($\equiv D(N,\mathbf{x,y})$) is $(\dim \mathcal{K})^{N-1}$. For
fixed $N$, $D(N,\mathbf{x,y})$ should be a monotonically decreasing function
of the lattice distance between $\mathbf{x}$ and $\mathbf{y}$. The
specification of this function is part of the definition of the quantum
space-time.

Most importantly, the time evolution operators in each sequence of Hilbert
spaces $\mathcal{H}_N (\mathbf{x})$ are constrained by the requirement that
they be compatible on all overlaps. This is such a complicated system of
constraints, that one might have despaired of finding a solution to it, if
it were not for the example of the $p=\rho$ universe discussed in the
previous section. We have yet to find a clue, which would help us to
construct an example of a universe that supports localized excitations.

For true scattering universes, the initial state is pure only as $N
\rightarrow \infty$. The Hilbert spaces of different observers must all
coincide in this limit. The S-matrix is expected to be unique and
mathematically well defined. The most interesting question for such
space-times is how one can express the constraints of compatibility of the
descriptions of different observers as equations for the S-matrix. We
conjecture that these equations will be generalizations of the usual
criteria of crossing symmetry and analyticity, and that, together with
unitarity, and a specification of the boundary geometry, they will
completely determine the S-matrix.

For Big Bang cosmologies, the construction is similar except that there is
an initial time slice, and all causal diamonds begin on that slice\footnote{%
Remember that we are working in a gauge-fixed formalism. This condition is
part of the gauge fixing.}.

\section{Discussion}

The phenomenological discussion of holographic cosmology presented in \cite%
{holcosm1and2} begins from a system close to the $p=\rho$ cosmology, but
requires inhomogeneous defects as input. We have treated these defects
heuristically as a network of spheres joined together in a ``tinker toy".
This was motivated by the observation that the Israel junction condition
applied to a single sphere of radiation or matter dominated cosmology
embedded in a $p=\rho$ background, requires the sphere to shrink in FRW
coordinates. The tinker toy is supposed to be the maximal entropy
configuration\footnote{%
which fits inside a given initial coordinate sphere. We will return to what
determines the initial size of this sphere.} for which this collapse does
not occur. To maximize the entropy we minimize the initial volume of the
normal region. The initial ratio of volumes is called $\epsilon $ and is
assumed small. We then argued that the volume of normal region, in equal
area slicing, grows relative to that of the $p=\rho$ region. Eventually, the
physical volume of the initial coordinate sphere is dominated by the normal
region. The $p=\rho$ regions are large black holes embedded in the normal
region. From this point on, the evolution can be treated by conventional
field theory methods, and we argued that it is plausible, if the low energy
degrees of freedom include an appropriate inflaton field, for the universe
to undergo a brief period of inflation. Depending on the value of $\epsilon$
(and another parameter which we cannot calculate), the fluctuations of the
microwave background can be generated either in the $p=\rho$ phase, or
during inflation. The two possibilities are incompatible with each other and
the experimental signatures of them are, in principle, distinguishable.

In order to put this cosmology on a mathematical basis, we have to find a
holographic description of a normal radiation dominated universe. Next we
must understand how the consistency conditions which we have discussed in
this paper, can be used to define an infinite hyperplanar boundary between a
normal phase and the dense black hole fluid. This would be the quantum
analog of the Israel junction condition. At this stage of development one
might hope to get a crude estimate of $\epsilon$. More detailed questions,
such as whether the fluctuations generated during the $p=\rho$ era have
Gaussian statistics, will probably require us to understand the more
complicated boundary of the tinker toy.

These problems seem hard, but before the present work we had despaired of
ever finding a solution to the consistency conditions for holographic
cosmology.

We want to end this paper with a metaphysical speculation. The Israel
junction condition applied to the large sphere inside of which the tinker
toy fits, would seem to require that that region collapse in coordinate
volume. One way to avoid this catastrophe would be to imagine that both the
initial black hole fluid, and the tinker toy had infinite extent in space.

There is a more attractive way out of this problem. If we try to embed a
(future) asymptotically de Sitter space into the $p=\rho$ fluid, we can
satisfy the Israel condition by matching the cosmological horizon to a
sphere of fixed physical size in the $p=\rho$ background. Now we imagine an
infinite $p=\rho$ background, littered with tinker toys of various sizes,
with the proviso that low energy physics inside each tinker toy universe is
compatible with eventual evolution to a stationary state of fixed positive
cosmological constant. From a global point of view, we would have a
collection of finite, asymptotically dS universes, embedded in an infinite,
flat $p=\rho$ background.

We can also understand the stability of this sort of cosmology from an
entropic point of view. We have advocated the $p=\rho$ cosmology as the most
entropic initial condition for the universe. In fact, in the more general
cosmology consisting of an infinite $p=\rho$ background, filled with a
collection of dS bubbles, any causal diamond which includes complete dS
bubbles, has the same number of states ``excited" as the pure $p=\rho$
fluid. It is only when we look at causal diamonds inside a dS bubble that we
find observers which observe less than the maximal amount of entropy. We
have argued that the most generic way for such low entropy regions to arise
is for the interior of the dS bubble to begin as a tinker toy embedded in a $%
p=\rho$ background. This then goes through a stage where the localized
entropy increases and is eventually followed by an AsdS stage where the
localized entropy is very small because everything has been swept out of the
observer's horizon.

Our notion of a generic state in an AsdS universe should be compared with
that of \cite{lennetal}. These authors organize the states according to the
eigenvalues of the static Hamiltonian. They then require that cosmological
evolution be viewed as a typical thermal fluctuation with certain constraints%
\footnote{%
The explicit model is a scalar field with an inflationary maximum and a dS
minimum with small c.c. . The typical cosmological fluctuation is one which
puts the scalar at the inflationary maximum.}. Among these constraints is
the anthropic principle. They then argue that a typical cosmology consistent
with these constraints will not look like the world we observe. From our
point of view, the choice of initial conditions made by these authors is not
the maximally entropic one for a local observer. They impose global
constraints on the states (thermality with respect to the static Hamiltonian
of the asymptotic future, and homogeneity over the inflationary horizon
size) at arbitrarily early times. On the contrary, in most early horizon
volumes we allow an absolutely random state to be acted on by a random
sequence of Hamiltonians. Certain horizon volumes, which contain parts of
the tinker toy, are somewhat more structured. In a previous paper we have
argued that these initial conditions have much more entropy than
inflationary ones. In our model, inflation only becomes possible in large
normal regions in which the black hole fluid has become dilute.

The $p=\rho$ universe with a distribution of AsdS bubbles is a model which
naturally provides us with an ensemble of universes with varying
cosmological constant. If we wish, we can apply the anthropic mode of
reasoning to this model. If the physics of a stable dS universe approaches a
limit as $\Lambda$ goes to zero, with the parameters which determine the
primordial density fluctuations and the dark matter density at the beginning
of the matter dominated era, both becoming independent of $\Lambda$ in the
limit, then Weinberg's anthropic argument for the value of the c.c. would
more or less explain the value that we see. At the very least, it explains
most of the ``fine tuning" that we find so disturbing.

We are of two minds as to the virtues of such a model. Much of our previous
work on the asymptotic dS universe simply postulates the cosmological
constant as an input, whose value will never have an explanation. The model
under discussion views that input as being determined by a very weak form of
the anthropic principle. We gain some degree of understanding\footnote{%
avoiding the introduction, by hand, of a huge integer, the number of
physical states, into our model of the world}, but at the expense of
introducing a large set of degrees of freedom which will never be observed.
Occam would surely complain!

On the positive side, one should compare this use of the anthropic principle
with others which have been contemplated in the literature. First of all, in
this model we imagine that all of the physics in a given tinker-toy universe
is completely determined by the value of a single parameter, the
cosmological constant. Thus, our model is required to calculate most
physical quantities successfully, from first principles. Only one parameter
is determined anthropically, and it is one for which the anthropic range is
quite narrow if everything else is fixed at its measured value. Secondly,
the anthropic argument we use is quite broad, and would apply to any form of
life whose existence depends on structures as complicated as galaxies. This
fixes the c.c. to be no larger than a factor of 100 times its observed
value. Even the more refined arguments of Vilenkin \cite{vil} , which reduce
this factor to something of order one, do not depend on crucial details of
nuclear physics or organic chemistry, as long as we view the c.c. as the
only parameter which varies among the different universes in our ensemble.

To summarize, we have described a well defined quantum mechanical model,
which obeys a plausible set of axioms for quantum cosmology. At large scales
it obeys scaling laws which are the same as those obeyed by a flat FRW
universe with equation of state $p=\rho$. The detailed dynamics of the model
realizes many of the properties of such a system that two of the authors
have proposed based on the intuitive idea of a dense black hole fluid. The
constants in the geometrical equations can mostly be fit by choices of
constants in the quantum mechanics, but we have found one constant which
seems to be determined unambiguously. Unfortunately it misses the geometric
prediction by a factor of $d - 2$.


\begin{appendix}

\section{Appendix}

\label{Appendix A}

\subsection{Intersection of causal diamonds}

In this sub-appendix we will determine the causal diamond
$\mathcal{D}_{M}$ with maximal FSB area, which is contained in the
intersection of two causal diamonds $\mathcal{D}_{1}$ and
$\mathcal{D}_{2}$ both starting at time $\eta _{1}$ and ending at time
$\eta _{2}$. We will solve the problem first in the simple case of
Minkowski spacetime and then in a general conformally flat spacetime.

So let's first consider Minkowski spacetime with dimension $d=4$
\begin{equation*}
ds^{2}=d\eta ^{2}-d\mathbf{x}^{2}
\end{equation*}
where we use the following notation $\mathbf{x}=(x,y,z)$ for the spatial
coordinates.

It will be clear in the following that identical considerations apply to
spacetimes of general dimension.

Given the two causal diamonds $\mathcal{D}_{1}$ and $\mathcal{D}_{2}$,
both starting at time $\eta _{1}$ and ending at time $\eta _{2}$, we will
indicate with $\mathcal{D}_{M}$ the maximal causal diamond belonging to
the intersection of $\mathcal{D}_{1}$ and $\mathcal{D}_{2}$
Fig.~\ref{3dcones} and Fig.~\ref{2dcausaldiamonds}. Let's indicate with
$\Sigma $ the spatial surface to which both the base sphere
$S_{\mathcal{D}_{1}}$ and $S_{\mathcal{D}_{2}}$ of $\mathcal{D}_{1}$ and
$\mathcal{D}_{2}$ belong. Let $\widetilde{S}$ be the maximal sphere that
fits into the intersection of
$S_{\mathcal{D}_{1}}$ and $S_{\mathcal{D}_{2}}$ . $\widetilde{S},S_{\mathcal{%
D}_{1}}$ and $S_{\mathcal{D}_{2}}$ are represented in Fig.~\ref{3dspheres}
and Fig.~\ref{2dspheres}. Furthermore let $(\eta _{i},\mathbf{x}_{i})$
$(\eta _{f},\mathbf{x}_{f})$ be the points of $\mathcal{D}_{1}$ $\cap $
$\mathcal{D}_{2}$ with the minimum and maximum values of $\eta $
Fig.~\ref{2dcausaldiamonds}.

It is obvious that the maximal causal diamond $\mathcal{D}_{M}$, belonging
to the intersection of two causal diamonds $\mathcal{D}_{1}$ and $\mathcal{D}%
_{2}$, must start at $(\eta _{i},\mathbf{x}_{i})$ end at $(\eta _{f},\mathbf{x%
}_{f})$ and have as base sphere $S_{\mathcal{D}_{M}}=\widetilde{S}$ .
In the $\eta ,x$ -plane Fig.~\ref{2dcausaldiamonds} we will indicate with $%
\Delta x$ the separation among the tips of $\mathcal{D}_{1}$ and $\mathcal{D}%
_{2}$ at time $\eta _{1}$.
The maximal causal diamond $\mathcal{D}_{M}$ will start at conformal time $%
\eta _{i}$ and end at conformal time $\eta _{f}$.

Denote by $r_{\mathcal{D}_{M}}$ the radius of the base sphere on $\mathcal{D}%
_{M}$ and with $h=\eta _{f}-\eta _{i}$ the height of the causal diamond $%
\mathcal{D}_{M}$ Fig.~\ref{2dcausaldiamonds} Fig.~\ref{maxdiamond1}.
Defining $h=2a$ we see from the pictures Fig.~\ref{2dcausaldiamonds} Fig.~%
\ref{maxdiamond1} we have $r_{\mathcal{D}_{M}}=a$. Furthermore we can see
inspecting Fig.~\ref{2dcausaldiamonds} that $\Delta x$
is given by%
\begin{equation*}
\eta _{i}-\eta _{1}=\frac{\Delta x}{2}
\end{equation*}
and so%
\begin{eqnarray*}
r_{\mathcal{D}_{M}} &=&a=\frac{1}{2}(\eta _{2}-\eta _{1})-(\eta _{i}-\eta
_{1}) \\
&=&\frac{1}{2}(\eta _{2}-\eta _{1})-\frac{\Delta x}{2}
\end{eqnarray*}

The quantities that we have determined, i.e. the radius of the base sphere $%
r_{\mathcal{D}_{M}}$, the height $h$ and the initial and final times $\eta
_{i},~\eta _{f}$, are all the parameters that describe the geometry of
$\mathcal{D}_{M}$.

We will now turn to the general problem of determining the maximal causal
diamond $\mathcal{D}_{M}$ in an FRW cosmology%
\begin{equation*}
ds^{2}=a^{2}(\eta )\left(d\eta ^{2}-d\mathbf{x}^{2}\right)
\end{equation*}
Since the space is conformally flat all the previous considerations
continue
to apply and the maximal causal diamond is still $\mathcal{D}_{M}$ Fig.~\ref%
{3dcones}, Fig.~\ref{3dspheres} and Fig.~\ref{2dspheres}. Moreover the
parameters that determine completely the geometry of $\mathcal{D}_{M}$ are,
as before, the radius of the base sphere $r_{\mathcal{D}_{M}}=\frac{1}{2}%
(\eta _{2}-\eta _{1})-\frac{\Delta x}{2}$, the height $h$ and the initial
and final times $\eta _{i},~\eta _{f}$.

Next we determine the sphere of maximal area (maximal sphere) on the
causal diamond $\mathcal{D}_{M}$ in an FRW\ cosmology
\begin{equation*}
ds^{2}=a^{2}(\eta )\left(d\eta ^{2}-d\mathbf{x}^{2}\right)
\end{equation*}
The area of a \emph{generic} 2-sphere $S$ of radius $r$, given by the
intersection of $\mathcal{D}_{M}$ and the spatial section at time $\eta $
Fig.~\ref{2dcausaldiamonds} Fig.~\ref{maxdiamond2}, is
\begin{equation}
A(\eta )=4\pi r^{2}a^{2}(\eta )  \label{area of 2 sphere on Dm}
\end{equation}
As mentioned before we want to determine the maximal area sphere $S_{M}$.

Assume that the spacetime contracts monotonically as we\ move toward the
past: ($a(\eta )$ decreases monotonically as $\eta $ goes to zero). Then
the maximal sphere is always in the upper half of the causal diamond Fig.~%
\ref{maxdiamond2} and its radius is
\begin{equation}
r=\eta _{f}-\eta  \label{radius}
\end{equation}

To determine the maximal sphere we have to maximize the area $A(\eta )$ in
the interval $(\eta _{f},\overline{\eta })$, where we defined $\overline{%
\eta }=\frac{1}{2}(\eta _{f}-\eta _{i})$.

The Friedmann's equations in conformal coordinate are
\begin{equation}
\frac{\dot{a}^{2}}{a^{2}}=\frac{8\pi \rho a^{2}}{3}-k
\label{Friedman equations}
\end{equation}
where $k=0$ for the conformally flat metric that we are considering. We
will assume as usual for an FRW\ cosmology that the matter content of the
universe is a perfect fluid with stress tensor%
\begin{equation*}
T_{a}^{b}=diag(-\rho ,p,p,p)
\end{equation*}
Assume that the pressure $p$ and energy density $\rho $ are related by the
equation of state%
\begin{equation*}
p=w\rho
\end{equation*}

With this ansatz for the matter content the Friedmann's Equations (\ref%
{Friedman equations}) can be solved and we find the conformal factor
\begin{equation}
a(\eta )=a_{0}\left(\frac{\eta }{q}\right)^{q},~~q=\frac{2}{1+3w}
\label{solution Friedman eqs}
\end{equation}

The extremum of area $A(\eta )$ is given by%
\begin{equation*}
\frac{dA(\eta )}{d\eta }=0
\end{equation*}%
using the Eq. (\ref{solution Friedman eqs}) for the conformal factor and the
expression (\ref{radius}) for the radius we find
\begin{equation*}
\widetilde{\eta }=\frac{q\eta _{f}}{1+q}
\end{equation*}
furthermore we have%
\begin{equation*}
\frac{d^{2}A(\widetilde{\eta })}{d\eta ^{2}}=-\frac{2(1+q)}{q}\left(\frac{%
\eta _{f}}{1+q}\right)^{2q}<0,~~\forall ~q,~\eta _{f}
\end{equation*}
showing that $\widetilde{\eta }$ is actually a maximum.

It is clear from Fig.~\ref{maxdiamond2}, that if
\begin{equation}
\widetilde{\eta }>\overline{\eta }=\frac{1}{2}(\eta _{f}-\eta _{i})
\label{base maximum condition 1}
\end{equation}
then the point where we have the maximal sphere is at $\eta
_{M}=\widetilde{\eta } $, otherwise the maximal sphere it is at $\eta
_{M}=\overline{\eta }$.

The condition given by Eq. (\ref{base maximum condition 1}) is equivalent to%
\begin{eqnarray}
\frac{q\eta _{f}}{1+q} &>&\frac{1}{2}(\eta _{f}-\eta _{i})  \notag \\
w &<&\frac{1}{3}\frac{\eta _{f}+3\eta _{i}}{\eta _{f}-\eta _{i}}
\label{base maximum condition w}
\end{eqnarray}
where the last quantity is clearly always greater than zero.

The previous condition (\ref{base maximum condition 1}) is always verified
for dust $w=0$ and for spacetime with a positive cosmological constant
$w=-1$, implying that in these case the maximal sphere is at $\eta _{M}=\widetilde{%
\eta }$ .

For a radiation dominated universe we have $w=\frac{1}{3}$ and the condition
(\ref{base maximum condition w}) becomes

\begin{eqnarray}
\frac{1}{3} &<&\frac{1}{3}\frac{\eta _{f}+3\eta _{i}}{\eta _{f}-\eta _{i}}
\notag \\
&\Rightarrow &  \notag \\
\eta _{i} &>&0  \label{base maximum condition radiation}
\end{eqnarray}
this is always true and so even in this case we have $\eta _{M}=\widetilde{%
\eta }$ .

The interesting case for the bulk of this paper is $w=1$. In this case if $%
\eta _{f}\gg 1$ (large enough causal diamonds) the condition (\ref{base
maximum condition w})%
\begin{equation}
w<\frac{1}{3}\frac{\eta _{f}+3\eta _{i}}{\eta _{f}-\eta _{i}}
\label{base maximum condition w=1}
\end{equation}
is never verified. As a consequence in this limiting case we always have $%
\eta _{M}=\overline{\eta }$, or in other words the maximal sphere
coincides with the base sphere of $\mathcal{D}_{M}$. This gives the area
formula we used in the text.

\subsection{Holographic relations in a general FRW cosmology}

In this sub-appendix, we want to show how the relation between area and
conformal time for a general FRW universe, filled with a combination of
perfect fluids, can be used to extract the equation of state. This indicates
that in a more general holographic cosmology, we can expect the formula for
the Hamiltonian as a function of the area to determine the background metric.

The metric for an FRW universe is
\begin{equation*}
ds^{2}=-dt^{2}+a^{2}(t)\left(\frac{dr^{2}}{1-kr^{2}}+r^{2}d\Omega ^{2}\right)
\end{equation*}
To analyze this problem it's more useful to work with conformal time $\eta
$ and comoving coordinate $\chi $%
\begin{equation*}
d\eta =\frac{dt}{a(t)},~~d\chi =\frac{dr}{\sqrt{1-kr^{2}}}
\end{equation*}%
\begin{equation*}
ds^{2}=a^{2}(\eta )\left(-d\eta ^{2}+d\chi ^{2}+f^{2}(\chi )d\Omega
^{2}\right)
\end{equation*}

Where as usual $k=-1,0,1$ and $f(\chi )=\sinh \chi ,\chi ,\sin \chi $
correspond to open, flat and closed universes, respectively.

We want to analyze a flat universe $f(\chi )=\chi $.
Consider the FRW universe with a big bang singularity, given any point \emph{%
p} in the space-time consider the backward light cone, it initially
expands and then starts contracting when we approach the singularity. Let
$B$ be the \emph{apparent horizon },i.e. the spatial surface with the
maximum area on the light cone. According to the covariant entropy bound,
the total number of degrees of freedom is bounded by the area of $B$
\begin{equation*}
N\leq\frac{A(B)}{2}
\end{equation*}
The apparent horizon is found geometrically as the sphere at which at
least one pair of lightsheets has zero expansion. The radius of the
apparent
horizon $\chi _{AH}(\eta )$, as a function of time, is given by the equation%
\begin{equation*}
\frac{\overset{.}{a}}{a}(\eta )=\pm \frac{f^{\prime }}{f}=\pm \frac{1}{\chi }
\end{equation*}
The proper area of the apparent horizon is given by%
\begin{equation*}
A_{AH}(\eta )=4\pi a^{2}(\eta )f^{2}[\chi _{AH}(\eta )]
\end{equation*}
In the case of a flat universe $f(\chi )=\chi $
\begin{equation*}
A_{AH}(\eta )=\frac{4\pi a^{2}(\eta )}{\frac{\dot{a}^{2}}{a^{2}}}
\end{equation*}
Using the Friedmann's equations (in conformal time)%
\begin{equation*}
\frac{\dot{a}^{2}}{a^{2}}=\frac{8\pi \rho a^{2}}{3}-k
\end{equation*}
with $k=0$, we have
\begin{equation*}
A_{AH}(\eta )=\frac{3}{2\rho (\eta )}
\end{equation*}

All these results are valid for cosmologies with a generic $\rho $. Thus,
the time dependence of the area of the apparent horizon determines the
time dependence of the energy density and vice versa.

We will first write everything as a function of cosmological scale factor,
so that the previous equation reads
\begin{equation*}
A_{AH}(a)=\frac{3}{2\rho (a)}
\end{equation*}
We want to determine $\rho (a)$ for a fluid with many components.
The equation of energy conservation for one fluid is%
\begin{equation*}
d\left(a^{3}(\rho +p)\right)=a^{3}dp
\end{equation*}
Assuming an equation of state%
\begin{equation*}
p=w\rho
\end{equation*}
this can be rewritten%
\begin{equation*}
\frac{d\rho }{da}+\alpha \frac{\rho }{a}=0
\end{equation*}
with%
\begin{equation*}
\alpha =3(1+w)
\end{equation*}
In general for many fluids we will have%
\begin{equation*}
\sum_{i}\frac{d\rho _{i}}{da}+\alpha _{i}\frac{\rho _{i}}{a}=0
\end{equation*}
with%
\begin{equation*}
\alpha _{i}=3(1+w_{i})
\end{equation*}

To keep things simple we will consider the case of two fluids, but the
results will be valid in the general case.

A general solution is given by%
\begin{equation*}
\rho =\rho _{1}+\rho _{2}
\end{equation*}
with%
\begin{equation*}
\frac{d\rho _{1}}{da}+\alpha _{1}\frac{\rho _{1}}{a}=f(a)
\end{equation*}%
\begin{equation*}
\frac{d\rho _{2}}{da}+\alpha _{2}\frac{\rho _{2}}{a}=-f(a)
\end{equation*}
so%
\begin{equation*}
\rho _{1}=C_{1}a^{-\alpha }+a^{-\alpha }\int_{a_{1}}^{a}d\bar{a}~\bar{a}%
^{\alpha }f(\bar{a})
\end{equation*}%
\begin{equation*}
\rho _{2}=C_{2}a^{-\alpha }-a^{-\alpha }\int_{a_{1}}^{a}d\bar{a}~\bar{a}%
^{\alpha }f(\bar{a})
\end{equation*}
with $a_{1}=a(\eta =1)$ and $C_{1}$ and $C_{2}$ integration constants.

In this context the form of the function $f(a)$ is not determined and so we
will consider $f(a)$ to be arbitrary. The function $f(a)$ describes how the
two fluids exchange energy and is determined by the dynamics of the system.

The area $A_{AH}$ will not depend on $w_{i}$ $\forall a$ iff%
\begin{eqnarray*}
\frac{\partial A_{AH}}{\partial w_{i}} &=&\frac{-3}{2\rho ^{2}}\frac{%
\partial \rho }{\partial w_{i}}=0,~\forall a \\
&\Longleftrightarrow & \\
\frac{\partial \rho }{\partial w_{i}} &=&0,~\forall a
\end{eqnarray*}
where we assumed $\rho \neq \infty $.

It turns out that there are no values of $C_{i},~a_{i},~f(a),~\alpha $ for
which%
\begin{equation*}
\frac{\partial \rho }{\partial w_{i}}=0,~\forall ~a
\end{equation*}
In fact considering for example$\frac{\partial \rho }{\partial w_{1}}$ we
have
\begin{eqnarray*}
\frac{\partial \rho }{\partial w_{1}} &=&\frac{\partial \rho
_{1}}{\partial
w_{1}} \\
&=&-3a^{-\alpha }\left(C_{1}\log (a)+\log (a)\int_{a_{1}}^{a}d\bar{a}~\bar{a}%
^{\alpha }f(\bar{a}) \right. \\
&&-\left. \left(\int_{a_{1}}^{a}d\bar{a}~\bar{a}^{\alpha }\log (\bar{a})f(\bar{a}%
)\right)\right)
\end{eqnarray*}
a necessary condition for this to be zero $\forall ~a$ is that the
derivative respect to $a$ is zero $\forall ~a$, where we assumed that
$a\neq 0$.
We have%
\begin{equation*}
\frac{\partial }{\partial a}\left(\frac{1}{-3a^{-\alpha }}\frac{\partial
\rho _{1}}{\partial w_{1}}\right)=\frac{1}{a}\left(C_{1}+\int_{a_{1}}^{a}d%
\bar{a}~\bar{a}^{\alpha }f(\bar{a})\right)
\end{equation*}
Assuming $a\neq \infty $ this can be zero $\forall ~a$ iff $C_{1}=f(a)=0$
but in this case we would have $\rho =0,~\forall ~a$.

As far as the dependence of $A_{AH}$ on the energy densities at some
initial time $\tilde{\rho}_{i}=\rho _{i}(\tilde{a})$, the area $A_{AH}$
will not depend on $\tilde{\rho}_{i}$ $\forall ~a$ iff%
\begin{eqnarray*}
\frac{\partial A_{AH}}{\partial \tilde{\rho}_{i}} &=&\frac{-3}{2\rho ^{2}}%
\frac{\partial \rho }{\partial \tilde{\rho}_{i}}=0,~\forall a \\
&\Longleftrightarrow & \\
\frac{\partial \rho }{\partial \tilde{\rho}_{i}} &=&0,~\forall a
\end{eqnarray*}
where we assumed $\rho \neq \infty $. It turns out that even in this case
there are no values of $C_{i},~a_{i},~f(a),~\alpha $ for which
\begin{equation*}
\frac{\partial \rho }{\partial \tilde{\rho}_{i}}=0,~\forall a
\end{equation*}
In fact%
\begin{eqnarray*}
\tilde{\rho}_{i} &=&\rho _{i}(\tilde{a})=C_{i}\tilde{a}^{-\alpha }+\tilde{a}%
^{-\alpha }\int_{a_{1}}^{\tilde{a}}d\bar{a}~\bar{a}^{\alpha }f(\bar{a}) \\
&\Longrightarrow & \\
C_{i} &=&\left(\frac{\tilde{\rho}_{i}}{\tilde{a}^{-\alpha }}-\int_{a_{1}}^{%
\tilde{a}}d\bar{a}~\bar{a}^{\alpha }f(\bar{a})\right)
\end{eqnarray*}
and%
\begin{eqnarray*}
\frac{\partial \rho }{\partial \tilde{\rho}_{i}} &=&\frac{\partial \rho _{i}%
}{\partial \tilde{\rho}_{i}}=\frac{\partial C_{i}}{\partial \tilde{\rho}_{i}}%
a^{-\alpha } \\
&=&\frac{a^{-\alpha }}{\tilde{a}^{-\alpha }}\neq 0,~\forall a
\end{eqnarray*}
always assuming that $a\neq 0$. Thus, we can always extract the parameters $%
w_i$ from the scale factor dependence of the energy density, and
consequently, from the scale factor dependence of the area of the apparent
horizon.

We now return to the problem of studying the dependence of $\rho $ as a
function of $\eta $ on the parameter $w_{i},~\tilde{\rho}_{i}$, which we
will now denote generically as $\beta _{i}$. We have%
\begin{equation*}
\rho =\rho \left(a(\eta ,\beta _{i}),\beta _{i}\right)
\end{equation*}
and so%
\begin{equation*}
\frac{\partial \rho }{\partial \beta _{i}}=\frac{\partial \rho }{\partial
\beta _{i}}+\frac{\partial \rho }{\partial a}\frac{\partial a}{\partial
\beta _{i}}
\end{equation*}
The problem is slightly more complicated but can still be solved exactly,
in fact the dependence of $a$ on $\beta _{i}$ can be found by solving the
Friedmann equations by quadrature%
\begin{equation*}
\frac{\dot{a}^{2}}{a^{2}}=\frac{8\pi \rho \left(a(\eta ),\beta
_{i}\right)a^{2}}{3}-k
\end{equation*}

We conclude that the component equations of state of an arbitrary
multi-component fluid, can be extracted from the dependence of the horizon
area on conformal time. In this derivation we have used the Friedmann
equation. In the quantum approach to cosmology, which we have discussed at
length in this paper, we believe that the replacement for the Friedmann
equation is the equation determining the $N$ dependence of the Hamiltonians $%
H_{N}(k,\mathbf{x})$. There are strong constraints on these Hamiltonians,
coming from the overlap conditions. We have found one solution of these
equations and argued that it corresponds to a $p=\rho $ FRW universe. We
conjecture that other solutions will also represent Big Bang cosmologies.

\subsection{Computation of $c_{e}$ from geometry and constant in front of $%
H_{N}$}

The Einstein equations in $d$ space-time dimensions are
\begin{equation*}
G_{\mu \nu }=2 \Omega _{d-1}G_{N}T_{\mu \nu }
\end{equation*}
where $\Omega _{d-1}$ is the surface of a sphere in $l=d-1$ spatial
dimensions. In $4$ dimension we recover the usual result
\begin{equation*}
G_{\mu \nu }=8\pi G_{N}~T_{\mu \nu }
\end{equation*}

Through a standard computation we recover the Friedmann's\ equation in $d$
dimensions%
\begin{equation*}
\left(\frac{\dot{a}}{a}\right)^{2}=\frac{8\left(\Omega _{d-1}G_{N}\right)%
}{d(1-d)}\rho
\end{equation*}

For a $p=\rho $ cosmology the expression for the energy density as a
function of the entropy density is%
\begin{equation*}
\rho =c_{e}{}^{2}\sigma ^{2}
\end{equation*}
substituting in the Friedmann's\ equation we have%
\begin{equation*}
\left(\frac{\dot{a}}{a}\right)^{2}=\frac{8\left(\Omega _{d-1}G_{N}\right)%
}{d(1-d)}\rho =\frac{c_{d}^{2}c_{e}^{2}\sigma _{0}^{2}}{a^{2(d-1)}}
\end{equation*}
the solution of the previous equation is
\begin{equation}
a(\eta )=a_{0}\eta ^{\frac{1}{d-2}}\left(a_{0}\frac{d-2}{d-1}\right)^{\frac{1%
}{d-2}}  \label{conformal factor}
\end{equation}
with%
\begin{equation}
a_{0}=\left(c_{d}c_{e}\sigma _{0}(d-1)\right)^{\frac{1}{d-1}}  \label{a 0}
\end{equation}%
\begin{equation*}
c_{d}=\sqrt{\frac{8\left(\Omega _{d-1}G_{N}\right)}{d(d-1)}}
\end{equation*}
in the following we will set%
\begin{equation*}
G_{N}=1
\end{equation*}
For a $p=\rho $ cosmology the value of the constant $c_{e}$ can be obtained
saturating the entropy bound.

We have for a causal diamond of maximal FSB area $A$
\begin{equation*}
A=a_{0}^{d-1}\left(\frac{d-2}{d-1}\right)\Omega _{d-1}\left(\frac{\eta }{2}%
\right)^{d-1}=4Nl_{s}=4\sigma _{0}\Omega _{d-1}\eta ^{d-1}\left(\frac{1}{d-1}%
\right)
\end{equation*}
substituting the equations (\ref{conformal factor}) and (\ref{a 0}) we find
\begin{equation*}
c_{e}=2^{d+1}\frac{1}{c_{d}}\frac{1}{(d-1)(d-2)}=2^{d+1}\frac{1}{\sqrt{\frac{%
8\left(\Omega _{d-1}\right)}{d(d-1)}}}\frac{1}{(d-1)(d-2)}
\end{equation*}
This expression for $c_{e}$ can be used to fix the constant in front of
the Hamiltonian $H_{N}$.

There do not seem to be any further consequences of requiring that our
quantum cosmology obey the equations of classical $p=\rho$ cosmology, not
just as scaling relations, but including the constants.   This only serves
to define Newton's constant, and the constant in front of our quantum
hamiltonian.   The one classical relation from which these constants scale
out is the relation between overlap areas.   Here we have a chance for a
numerical triumph, but our current definitions miss by a factor of $d -
2$.


\acknowledgments TB and WF would like to thank the KITP for it's hospitality
during the "String Theory and Cosmology" workshop when parts of this paper
were begun. The research of W. Fischler is based upon work supported by the
National Science Foundation under Grant No. 0071512. The research of T.
Banks and L. Mannelli was supported in part by DOE grant number
DE-FG03-92ER40689.



\newpage

\section{Figures}

\begin{figure}[h]
\includegraphics[bb=80 20 465
450,scale=1]{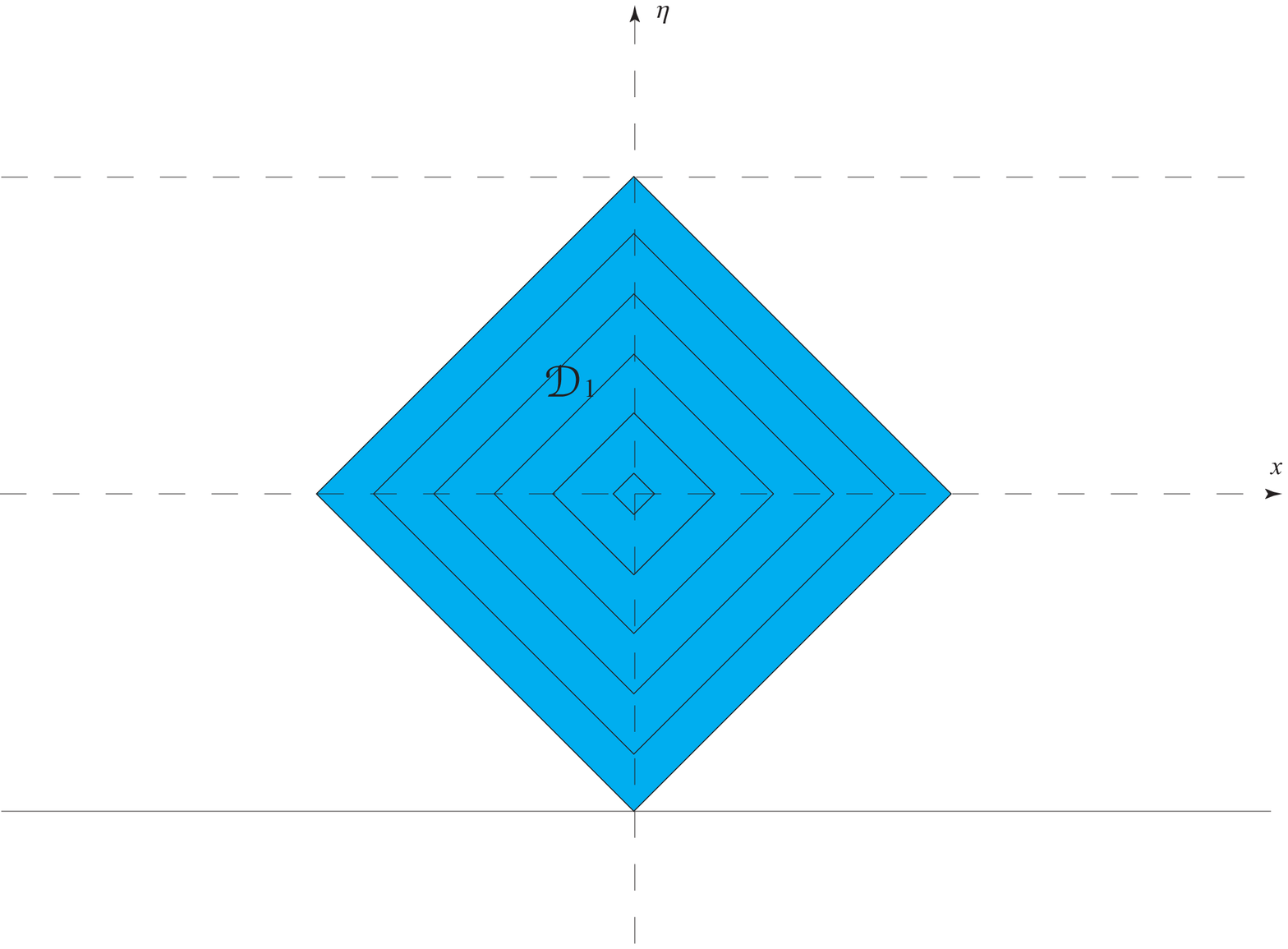} \caption{Nested causal diamonds defining an
observer in a time symmetric space-time} \label{nesteddiamonds}
\end{figure}

\begin{figure}[h]
\includegraphics[bb=80 20 465
450,scale=1]{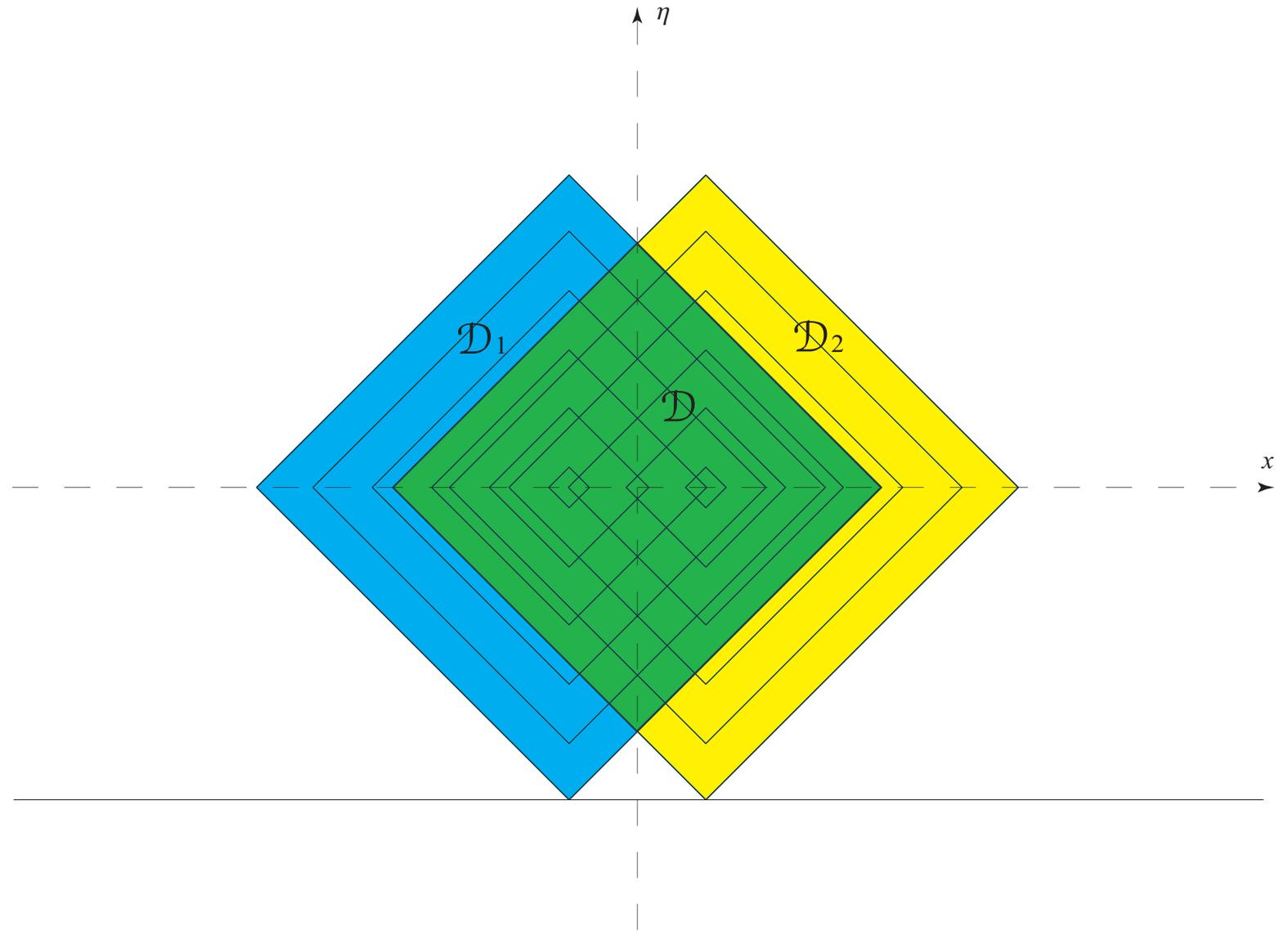} \caption{Nested causal diamonds
defining a nearest neighbors pair of observers in a time symmetric
space-time} \label{intersectionnesteddiamonds}
\end{figure}

\begin{figure}[h]
\includegraphics[bb=135 0 530 450,scale=1]{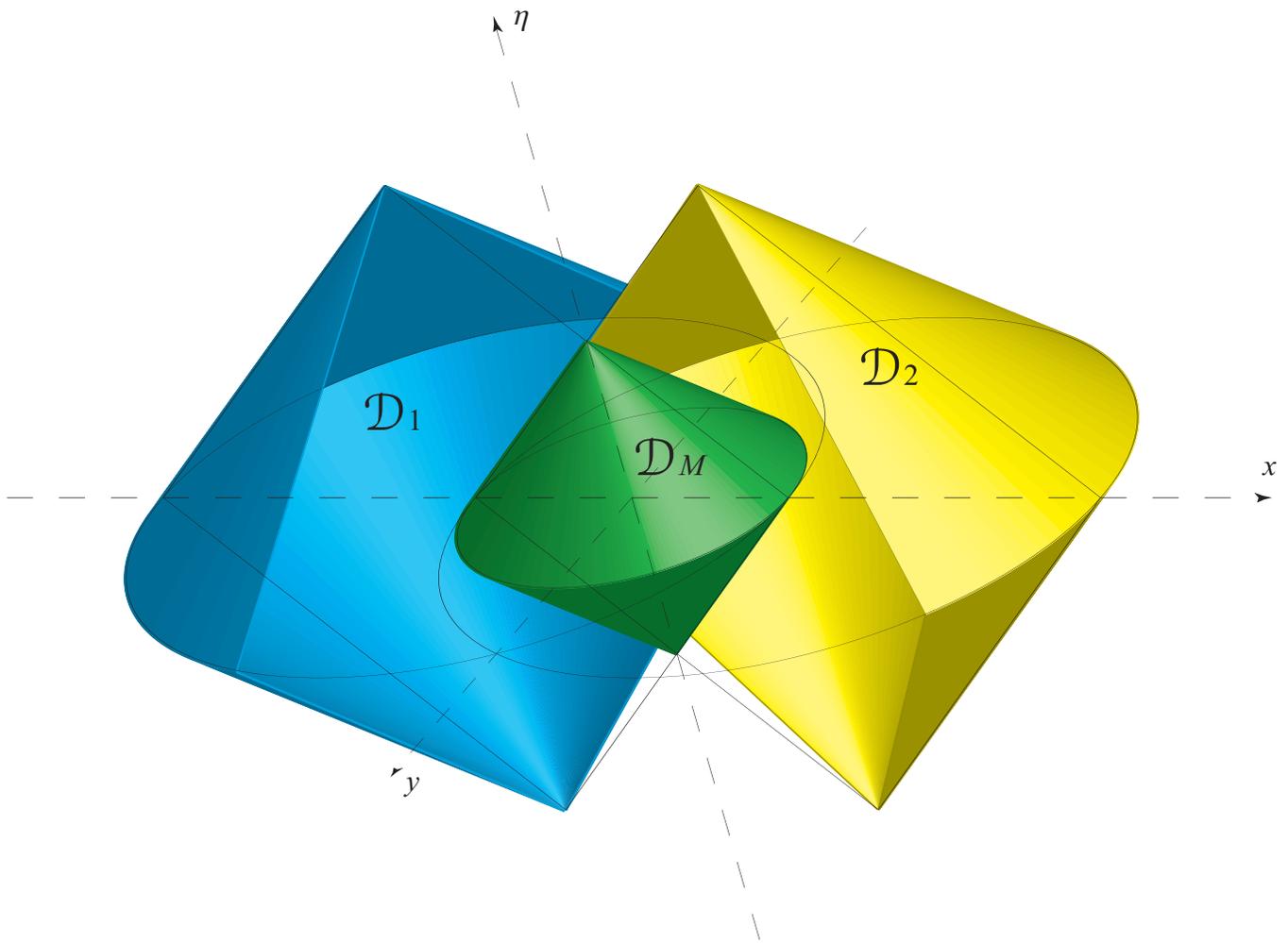}
\caption{The two causal diamonds $\mathcal{D}_{1}$ and $\mathcal{D}_{2}$
($z
$ spatial coordinate suppressed) and the maximal causal diamond $\mathcal{D}%
_{M}$ that fits in the intersection $\mathcal{D}_{1}$ $\cap $ $\mathcal{D}%
_{2}$. The picture is valid for Minkowski spacetime and more generaly
conformally flat spacetimes. } \label{3dcones}
\end{figure}

\begin{figure}[h]
\includegraphics[bb=135 0 530 450,scale=1]{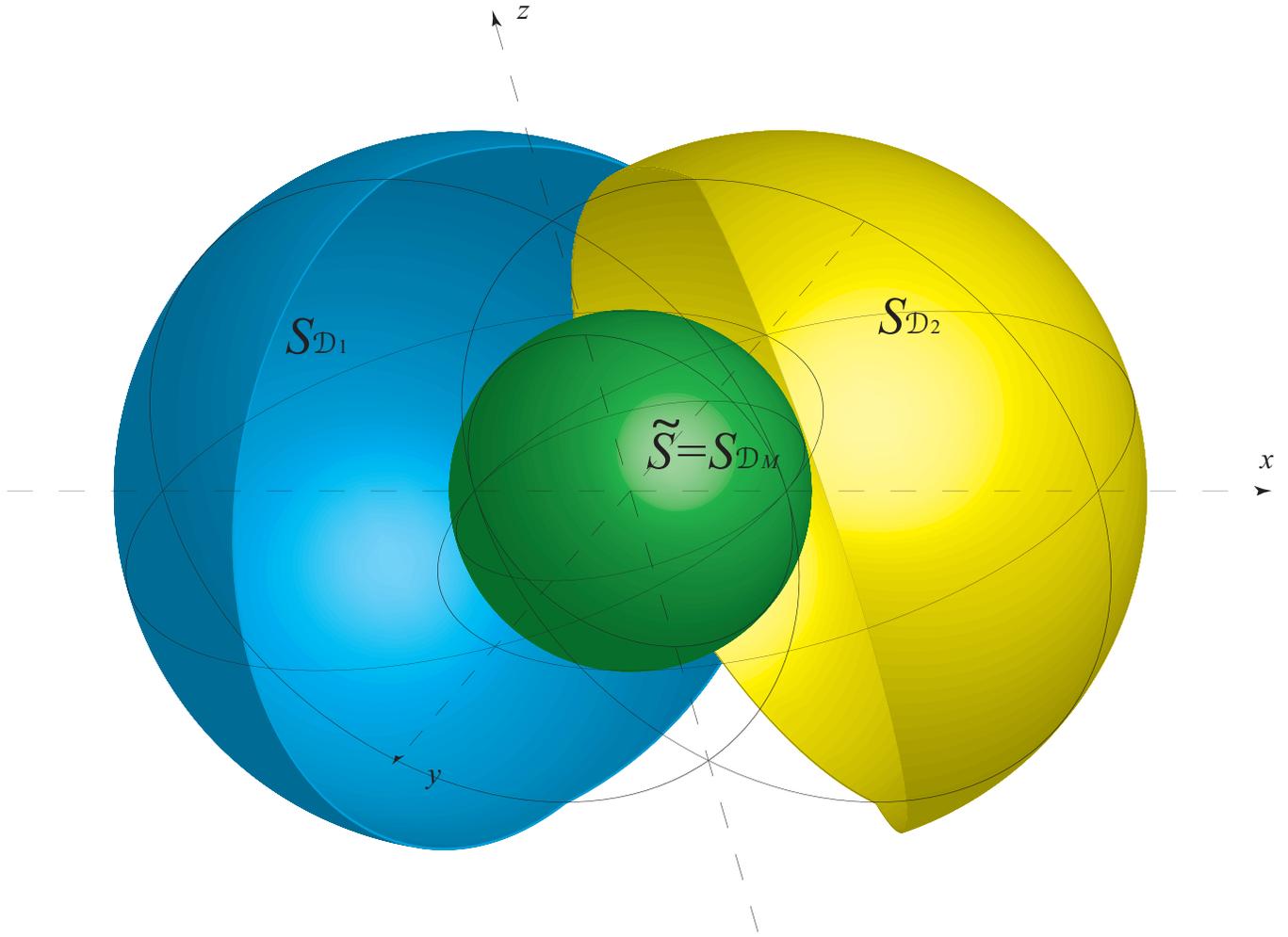}
\caption{The base spheres $S_{\mathcal{D}_{1}}$ and
$S_{\mathcal{D}_{2}}$of
the two causal diamonds $\mathcal{D}_{1}$ and $\mathcal{D}_{2}$ (Fig.~%
\protect\ref{3dcones}) $\widetilde{S}$ is the maximal sphere belonging to
the intersection of $S_{\mathcal{D}_{1}}$ and $S_{\mathcal{D}_{2}}$. $%
\widetilde{S}$ coincide with $S_{\mathcal{D}_{M}}$ the base sphere of $%
\mathcal{D}_{M}$.} \label{3dspheres}
\end{figure}

\begin{figure}[h]
\includegraphics[bb=80 0 505 450,scale=1]{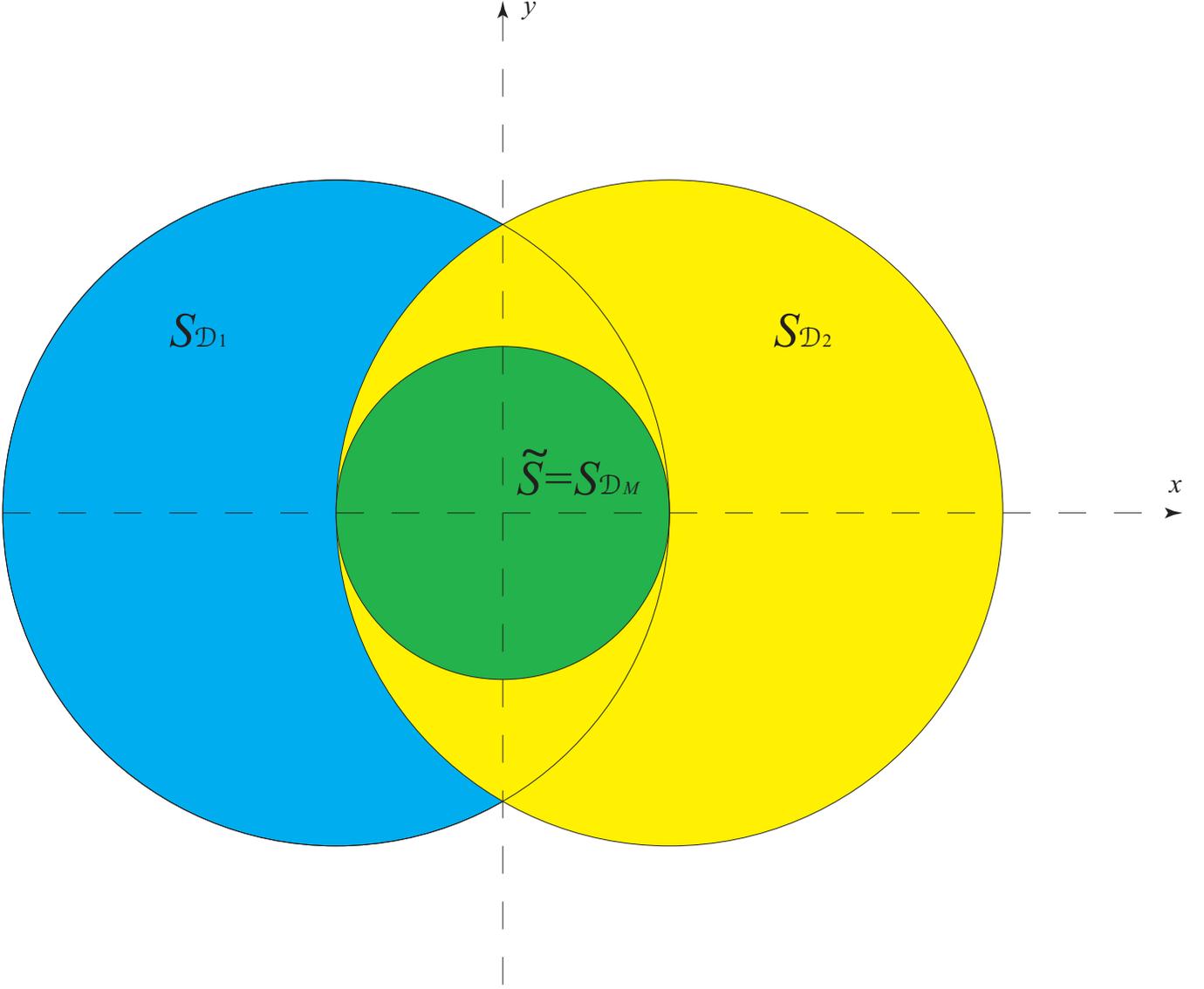}
\caption{The base spheres $S_{\mathcal{D}_{1}}$ and $S_{\mathcal{D}_{2}}$
of the two causal diamonds $\mathcal{D}_{1}$ and $\mathcal{D}_{2}$ (Fig.~%
\protect\ref{3dcones}), coordinate $z$ suppressed.} \label{2dspheres}
\end{figure}

\begin{figure}[h]
\includegraphics[bb=80 0 505 450,scale=1]{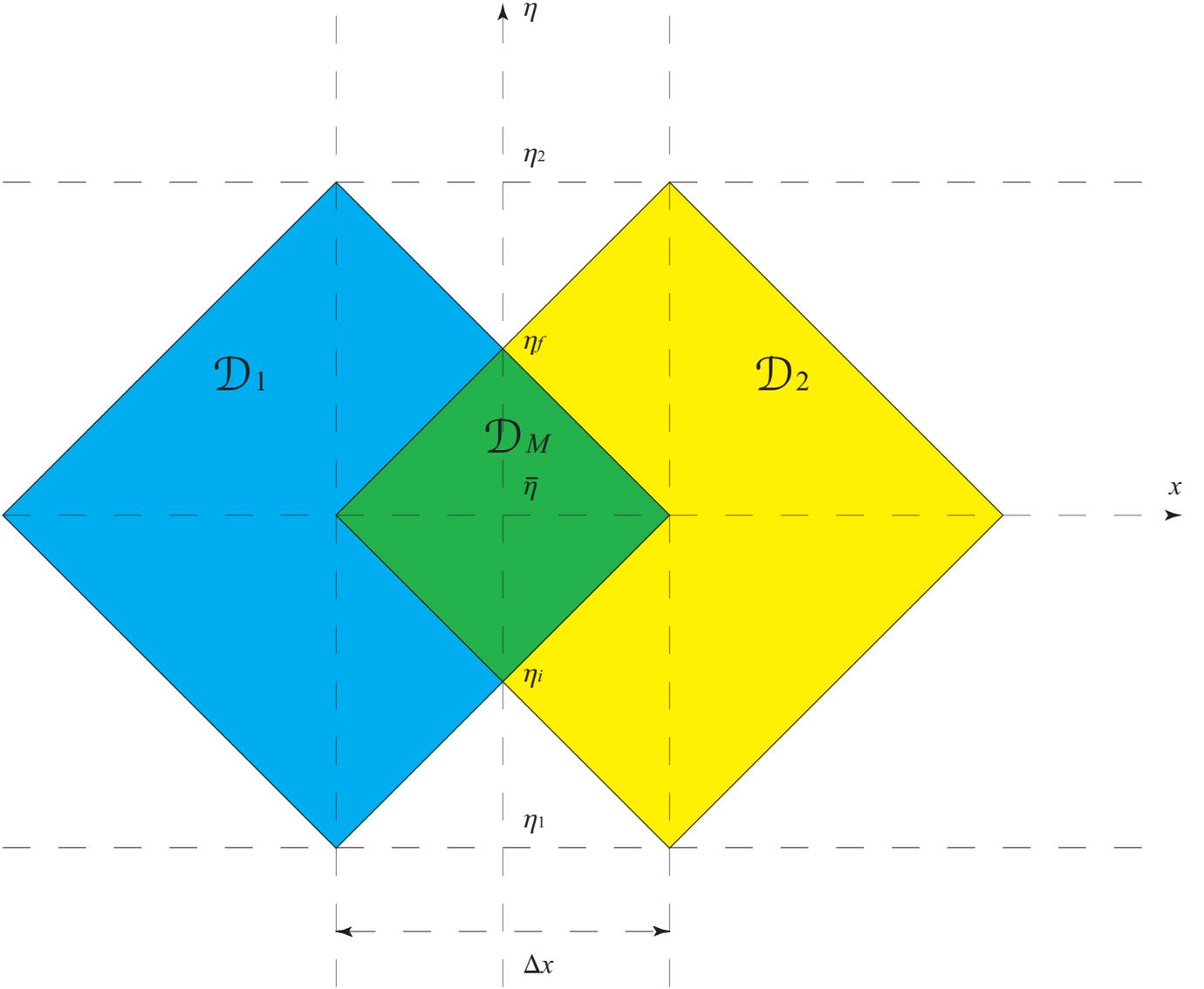}
\caption{The two causal diamonds $\mathcal{D}_{1}$ and $\mathcal{D}_{2}$ ($%
z,~y$ spatial coordinates suppressed) and the maximal causal diamond $%
\mathcal{D}_{M}$ that fits in the intersection $\mathcal{D}_{1}$ $\cap $ $%
\mathcal{D}_{2}$. $\protect\eta _{i},$ $\protect\eta _{f}$ are the points
of $\mathcal{D}_{1}$ $\cap $ $\mathcal{D}_{2}$with the minimum and maximum
values of $\protect\eta $. $\Delta x$ is the separation among the tips of $%
\mathcal{D}_{1}$ and $\mathcal{D}_{2}$ at time $\protect\eta _{1}$.}
\label{2dcausaldiamonds}
\end{figure}

\begin{figure}[h]
\includegraphics[bb=0 20 405 450,scale=1]{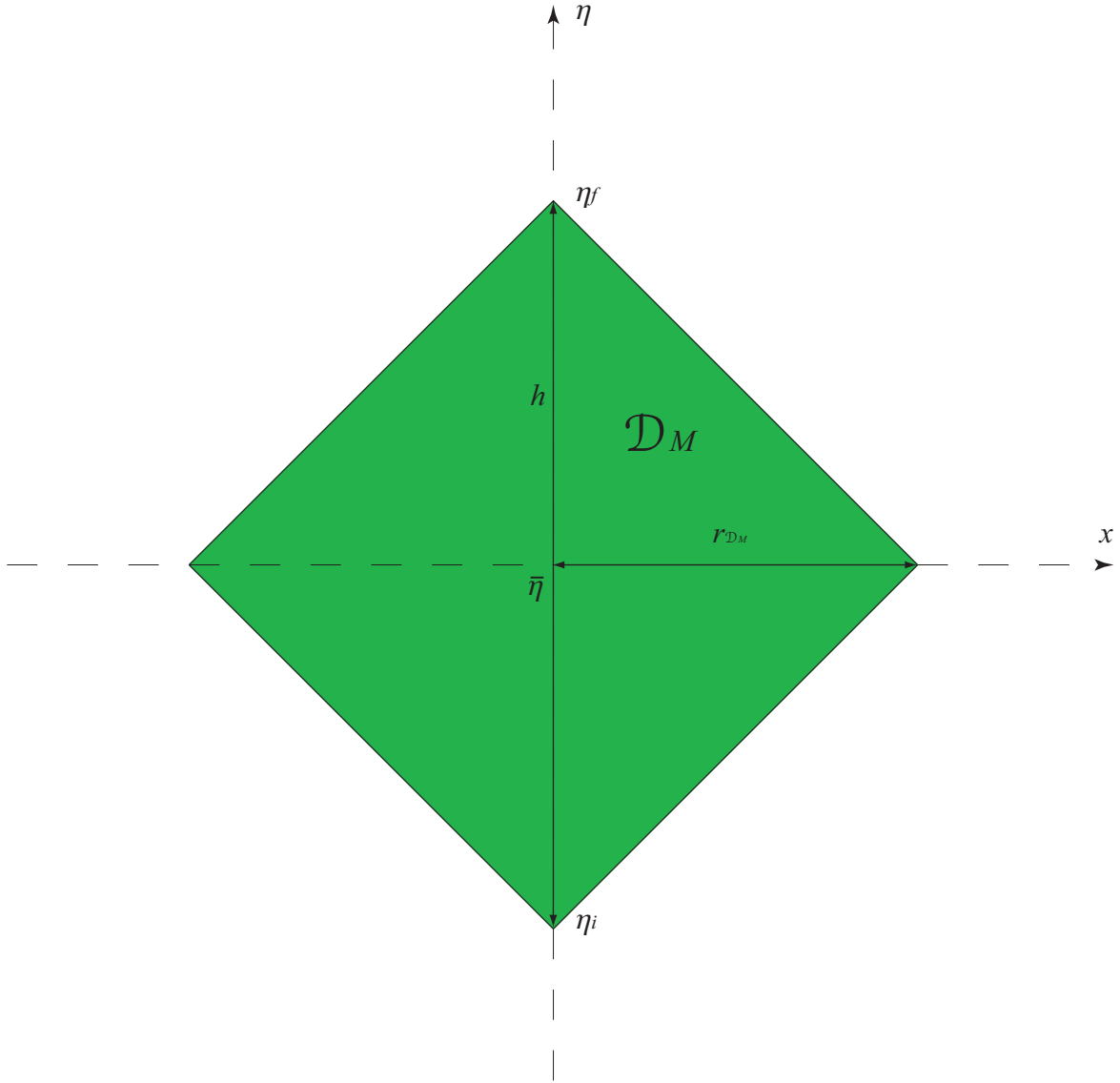}
\caption{Detail of the the maximal causal diamond $\mathcal{D}_{M}$. $h=%
\protect\eta _{f}-\protect\eta _{i}$ is the height of $\mathcal{D}_{M}$ and $%
r_{\mathcal{D}_{M}}$ is the radius of the base sphere on
$\mathcal{D}_{M}$. } \label{maxdiamond1}
\end{figure}

\begin{figure}[h]
\includegraphics[bb=0 20 405 450,scale=1]{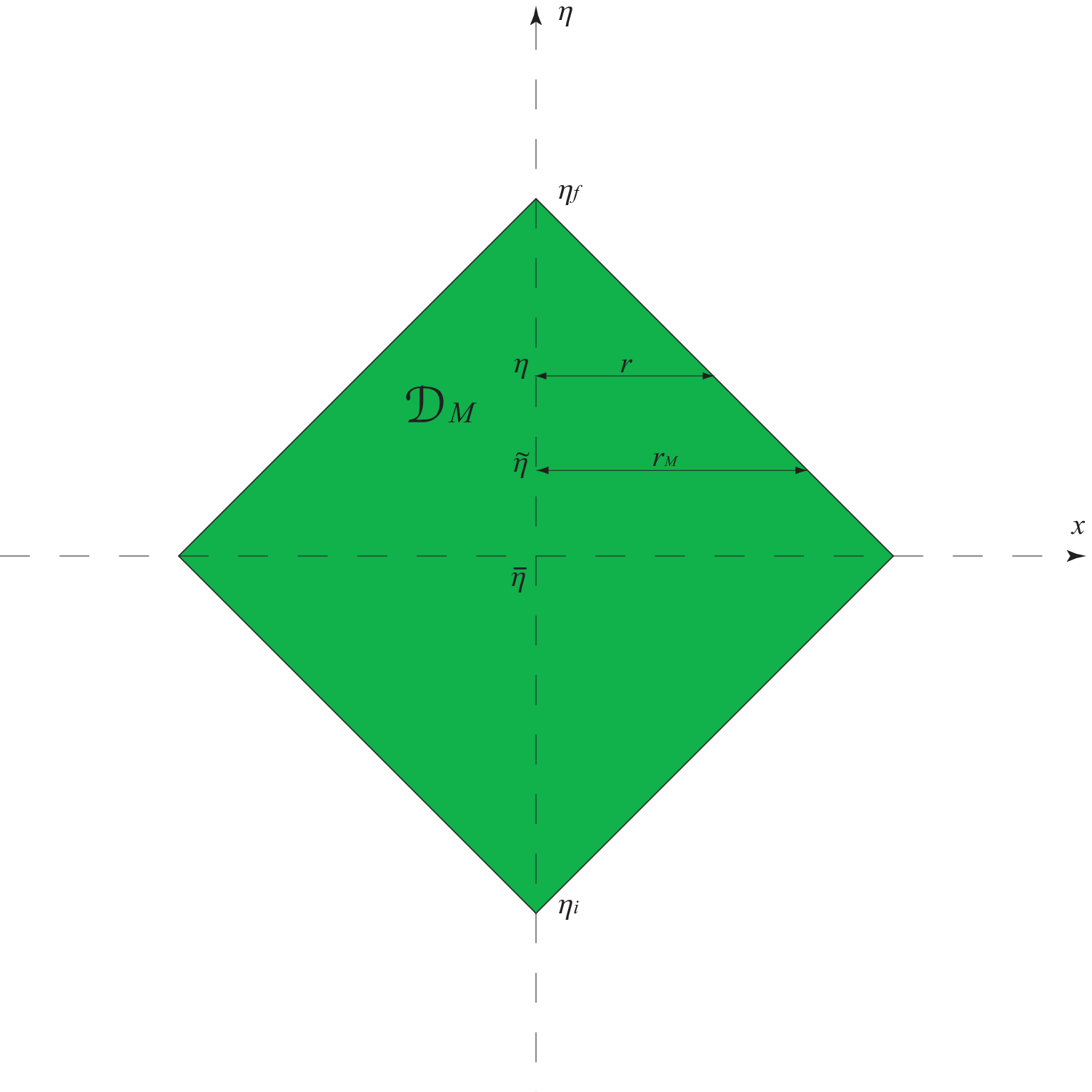}
\caption{Detail of the the maximal causal diamond $\mathcal{D}_{M}$. $r$
is the radius of a generical sphere on $\mathcal{D}_{M}$, $r_{M}$ is the
radius of the sphere of maximal area on $\mathcal{D}_{M}$ i.e. the maximal
sphere.} \label{maxdiamond2}
\end{figure}

\end{appendix}


\newpage

\begin{thebibliography}{99}


\bibitem{holocosmo}
T.~Banks and W.~Fischler, \textit{An holographic cosmology}
, hep-th/0111142 

T.~Banks and W.~Fischler, \textit{Holographic cosmology 3.0} ,
hep-th/0310288 

T.~Banks and W.~Fischler, \textit{Holographic cosmology}
, hep-th/0405200 

\bibitem{fsb} W.~Fischler, L.~Susskind,  \textit{Holography and Cosmology,}
hep-th/9806039;

R. Bousso, \textit{Holography in general space-times,} JHEP 9906, 028
(1999) hep-th9906022

E.~Flanagan, D.~Marolf, R.~Wald, {it Proof of classical versions of the
Bousso entropy bound qnd of the generalized second law,} hep-th/9908070


\bibitem{g} G.~Veneziano, Phys. Lett. \textbf{B454}, 22 (1999)

R.~Easther, D.~Lowe, Phys. Rev. Lett. \textbf{82}, 4967(1999)

D.~Bak, S.~Rey, \textit{Cosmic holography,} hep-th/9902173

N.~Kaloper, A.~Linde, Phys. Rev. \textbf{D60} 103509(1999)

R.~Brustein, G.~Veneziano, Phys. Rev. Lett.\textbf{84} 5695(2000)


\bibitem{susyholo3.0} T.~Banks, \textit{SUSY and the holographic screens,}
hep-th/0305163,

T.~Banks, W.~Fischler, \textit{Holographic Cosmology 3.0}, hep-th/0310288


\bibitem{BKL} V.A.~Belinskii, I.M.~Khalatnikov, E.M.~Lifshitz, \textit{%
Oscillatory approach to a singular point in the relativistic cosmology}
Adv.in Phys.,\textbf{19}, 525(1970)

\bibitem{holcosm1and2} T.~Banks, W.~Fischler, \textit{An Holographic
Cosmology}, hep-th/0111142

T.~Banks, W.~Fischler, \textit{Holographic Cosmology 3.0}, hep-th/0310288


\bibitem{cp} E.~Cartan, \textit{The Theory of Spinors,} MIT Press, Cambridge
MA, reprinted from the original French edition, Hermann, Paris, 1966

\bibitem{rubetal} V.~G.~Lapchinsky and V.~A.~Rubakov, \textit{Quantum
Gravitation: Quantization Of The Friedmann Model. (In Russian),} Teor.
Mat. Fiz. \textbf{33}, 364 (1977).

T.~Banks, W.~Fischler and L.~Susskind, \textit{Quantum Cosmology In
(2+1)-Dimensions And (3+1)-Dimensions,} Nucl. Phys. B \textbf{262}, 159
(1985).

T.~Banks, \textit{T C P, Quantum Gravity, The Cosmological Constant And
All That..,} Nucl. Phys. B \textbf{249}, 332 (1985).

V.~A.~Rubakov, \textit{On The Third Quantization And The Cosmological
Constant,} Phys.\ Lett.\ B \textbf{214}, 503 (1988).

G.~V.~Lavrelashvili, V.~A.~Rubakov and P.~G.~Tinyakov, \textit{Third
Quantization Of Gravity And The Cosmological Constant Problem,} In Moscow
1990, Proceedings, Quantum gravity (QC178:S4:1990), 27-43


\bibitem{raph} R. Bousso, \textit{Holography in general space-times,} JHEP
\textbf{9906}, 028 (1999) hep-th9906022


\bibitem{tbmillmcosmoholcosm} T.~Banks \textit{Supersymmetry and Space Time}
Talk  given at the Strings at the Millenium Conference, CalTech-USC,
January 5, 2000.

T.~Banks, W.~Fischler, \textit{M-theory observables for cosmological
spacetimes,} hep-th/0102077.

T.~Banks, W.~Fischler, \textit{An Holographic Cosmology}, hep-th/0111142

T.~Banks, W.~Fischler, \textit{Holographic Cosmology 3.0}, hep-th/0310288


\bibitem{cartpen} E.~Cartan, \textit{The Theory of Spinors,} MIT Press,
Cambridge MA, reprinted from the original French edition, Hermann, Paris,
1966

C.~W.~Misner, K.~S.~Thorne, J.~A.~Wheeler, \textit{Gravitation,} 41.9, p.
1157, W.H. Freeman, san Francisco, 1973, and references therein

\bibitem{sorokin} D.~P.~Sorokin, \textit{Supermembranes and Superembeddings,}
Phys. Rept. \textbf{329}, 1 (2002), hep-th/9906142


\bibitem{tbholosugra} T.~Banks, in preparation

\bibitem{connes} A.~Connes, Noncommutative Geometry, Academic Press Inc.,
San diego, CA 1994

\bibitem{KapluWein} F.~J.~Dyson, J.Math.Phys. \textbf{3}, 140 (1962),
\textbf{3} 1199 (1962)

V.~Kaplunovsky, M.~Weinstein, \textit{Space-Time: Arena or Illusion?}
Phys. Rev. \textbf{D31}, 1879 (1985)

\bibitem{mcosmo} T.~Banks, W.~Fischler, \textit{M-theory observables for
cosmological spacetimes,} hep-th/0102077.


\bibitem{nightmare} T.~Banks, W.~Fischler and S.~Paban, \textit{Recurrent
nightmares?: Measurement theory in de Sitter space.}  JHEP \textbf{0212},
062 (2002), hep-th/0210160.


\bibitem{bhcomplementarity} L.~Susskind, L.~Thorlacius, J.~Uglum, {it The
stretched horizon and black hole complementarity,} Phys. Rev.\textbf{D48},
3743 (1993), hep-th/9306069

G.~'t~Hooft, C.~R.~Stevens, B.~F.~Whiting,{Black hole evaporation without
information loss,} Class. Quantum Grav. \textbf{11}, 621 (19994),
gr-qc/9310006

G.~'t~Hooft,\textit{Quantum information and information loss in general
relativity,} in Quantum Coherence and Decoherence, K. Fujikawa, Y.A. Ono,
Eds,. Elsevier Science B.V., \textbf{273} (1996), gr-qc/9509050

G.~'t~Hooft, \textit{The scattering matrix approach for the quantum black
hole: an overview,} J. Mod. Phys. \textbf{A 11}, 4623 (1996), gr-qc/
9607022


\bibitem{vil} A.~Vilenkin, \textit{Predictions from quantum cosmology,}
Phys. Rev. Lett. \textbf{74}, 846 (1995), gr-qc/9406010


\bibitem{lennetal}
L.~Dyson, M.~Kleban and L.~Susskind, \textit{Disturbing implications of a
cosmological constant} JHEP {\bf 0210}, 011 (2002) hep-th/0208013

\end{thebibliography}
\end{document}